\documentclass[twocolumn]{aastex631}

\begin{document}

\title{X-ray and radio polarimetry of the neutron star low mass X-ray binary GX~13$+$1}

\author{Unnati Kashyap}
\affiliation{Department of Physics and Astronomy, Texas Tech University, Lubbock, TX 79409-1051, USA} 

\author{Thomas J. Maccarone}
\affiliation{Department of Physics and Astronomy, Texas Tech University, Lubbock, TX 79409-1051, USA} 

%\collaboration{20}{(AAS Journals Data Editors)}

\author{Eliot C. Pattie}
\affiliation{Department of Physics and Astronomy, Texas Tech University, Lubbock, TX 79409-1051, USA}

\author{Mason Ng}
\affiliation{Department of Physics, McGill University, 3600 rue University, Montr\'{e}al, QC H3A 2T8, Canada}
\affiliation{Trottier Space Institute, McGill University, 3550 rue University, Montr\'{e}al, QC H3A 2A7, Canada}

\author{Swati Ravi}
\affiliation{MIT Kavli Institute for Astrophysics and Space Research, Massachusetts Institute of Technology, Cambridge, MA 02139, USA}

\author{Herman L. Marshall}
\affiliation{MIT Kavli Institute for Astrophysics and Space Research, Massachusetts Institute of Technology, Cambridge, MA 02139, USA}



\begin{abstract}
We report the X-ray and radio polarization study of the neutron star (NS) low-mass X-ray binary (LMXB) GX~13$+$1 using the Imaging X-ray Polarimetry Explorer (IXPE) and Very Large Array (VLA). Simultaneous Neutron Star Interior Composition Explorer (NICER) observations show that the source was in parts of the Z state during our IXPE observations, exhibiting moderate changes in the hardness intensity diagram. The source exhibits X-ray dips in the light curve along with hints of polarization swings between the dip and non-dip states. The X-ray spectro-polarimetry results suggest a source geometry comprising an accretion disk component representing the softer disk emission, along with a blackbody representing the harder emission from the boundary layer (BL) or a spreading layer (SL). We investigate the geometry of GX 13+1 by considering our X-ray and radio polarization findings.

\end{abstract}



\section{Introduction} \label{sec:intro}

Low-mass X-ray binaries (LMXBs) hosting a weakly magnetized ($10^{7}-10^{9}$ G) neutron star (WMNS) are among the brightest known X-ray sources. They are composed of a neutron star (NS) that accretes matter from a low-mass companion star via Roche-lobe overflow. Based on the shape they trace out in the Colour-Colour diagram (CCD)/hardness intensity diagram (HID), they have been divided into two main classes: The atoll and Z sources \citep{1989A&A...225...79H}. The atoll sources (luminosity $\sim$ $0.01-0.1 L_{edd}$) trace out a well-defined banana (bright atolls) and island state  (LHS atolls) in the HIDs \citep{1995foap.conf..213V}. On the contrary, the Z-sources with higher brightness (luminosity $\sim L_{edd}$) trace out Z-shaped tracks in the HIDs consisting of three branches, called the horizontal branch (HB), the normal branch (NB), and the flaring branch (FB) \citep{1989A&A...225...79H}. 

The WMNS sources are known to show an evolution in the source spectral and timing properties with varying spectral states or accretion rates \citep{2001AdSpR..28..307B,2004astro.ph.10551V,2007A&ARv..15....1D,2007ApJ...667.1073L}. The origin of the different X-ray emission components in the case of WMNS sources is still a matter of debate due to the spectral modeling degeneracy. The X-ray spectrum is represented by a Comptonization component, along with either a direct black-body (BB) emission attributed to the NS surface (Western Model, \cite{1988ApJ...324..363W}), or a direct contribution from the accretion disk (Eastern Model, \cite{1989PASJ...41...97M}). The Western model assumes that the hard component originates from the Comptonization of disk photons by the hot energetic electrons of the corona. In the Eastern model, however, the hard Comptonized emission is attributed to a boundary layer (BL) between the disk and the NS or a vertically extended spreading layer (SL) around the NS.

 As the sources transition through different X-ray spectral states, an evolution in the radio jet emission is also observed \citep{2006csxs.book..381F}.  Studying the X-ray polarization angle (PA) and its alignments with the radio jet position angle may provide a significant
understanding of the accretion geometry, with the jets providing a reference for the accretion disk orientation in WMNS.
Recent studies report alignment of the X-ray PA with the position angle of the radio jet in the case of Cyg X-2 \citep{2023MNRAS.519.3681F}, GX~5$-$1 (\cite{2024AA...684A.137F}, Pattie et al., in prep), and GX~17$+$2 (Kashyap et al., APJ, submitted). However, Sco X-1 shows a misalignment between the X-ray PA and the position angle of the radio jet \citep{2024ApJ...960L..11L}. Such misalignments are often associated with possibilities such as a change in the geometry of the X-ray emitting regions as the source evolves over time or the estimated polarization representing an overall polarized emission possibly coming from a superposition of different X-ray emission components. However, further investigations with more sensitive X-ray and radio observations of WMNS sources are crucial to disentangle their highly debated accretion geometry.

GX~13$+$1 is a peculiar source exhibiting both atoll and Z-source behavior. GX~13$+$1, located at a distance of  7 $\pm$ 1 kpc, with a late-evolved K5 III giant as a companion \citep{2002ApJ...570..793B} is the longest orbital period NS LMXB. It is as bright as a Z source \citep{2014A&A...564A..62D} with persistent radio emission \citep{1986ApJ...310..172G}, but it follows an atoll track in the HID/ CCD \citep{2003A&A...406..221S}. However, recent observations also show a possible Z track in the CCD \citep{2023MNRAS.522.3367S,2018ApJ...861...26A}. The source is known to exhibit a complex X-ray spectrum with the presence of reflection and absorption features caused by the interaction of the radiation in the wind above the disk, indicating a complicated structure of the wind \citep{2018ApJ...861...26A}. \cite{2020MNRAS.497.4970T} using the Fe XXV and XXVI absorption lines seen in the highest resolution Chandra third-order HETGS data gave constraints on the radial and azimuthal velocity of the wind. Along with the dips associated with the orbital movement, the source exhibits short dips occasionally interpreted as an indication of the high source inclination (60$\degr$ - 70$\degr$) \citep{2010AIPC.1248..153D,2014A&A...561A..99I} due to the interaction of the emission of the NS with the mostly neutral accretion bulge located in the outer parts of the disk \citep{2014A&A...564A..62D}.  GX~13$+$1 is one of the most interesting NS LMXB, exhibiting strong wind and switching polarization properties along with the presence of frequent dips in the light curve \citep{2024A&A...688A.217B,2024A&A...688A.170B}.

In this paper, we report IXPE observations of the WMNS GX~13$+$1 performed from 20 April, 2024 to 23 April, 2024, along with semi-simultaneous NICER observations. Our radio observations with the Very Large Array (VLA), along with the X-ray polarization results put constraints on the accretion geometry of GX~13$+$1.

\section{observations} \label{sec:style}
\label{obs}
\begin{figure}
\centering
\includegraphics[width=0.5\textwidth]{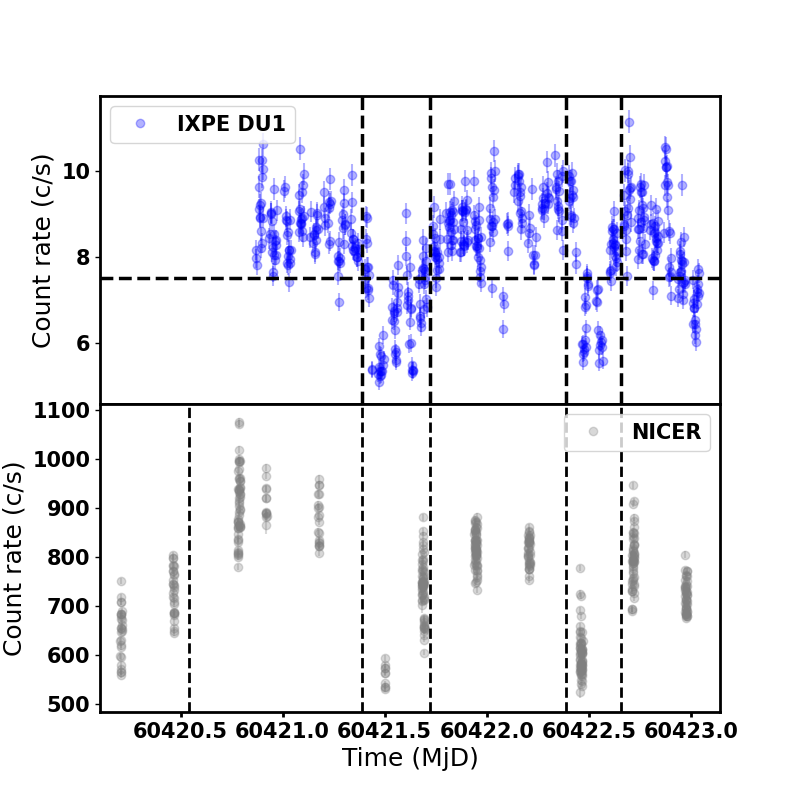}
\caption{Upper panel: IXPE (2-8 keV) light curve of GX~13$+$1. Lower Panel: NICER (0.5-10 keV) light curve of GX~13$+$1.
Time bins of 150 s (IXPE) and 8 s (NICER) are used.}
\label{lc}
\end{figure}

\begin{figure}
\centering
\includegraphics[width=0.5\textwidth]{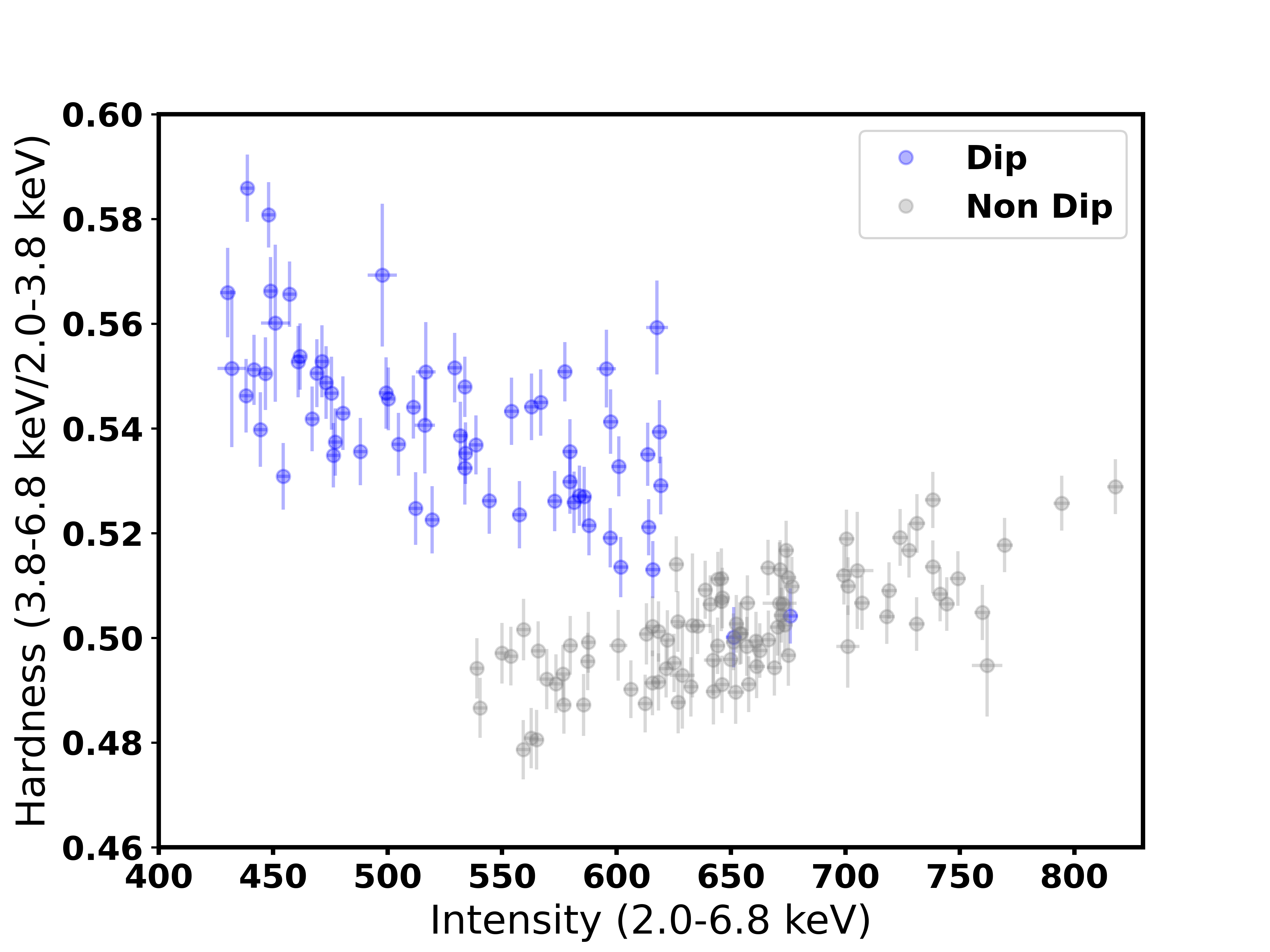}
\caption{Hardness Intensity diagram showing NICER observations of GX~13$+$1. Time bins of 64 s are used.}
\label{hid}
\end{figure}

\begin{table*}
\centering

\caption{ {\em IXPE}, {\em NICER}, and the {\em VLA} Observations of GX~13$+$1 (see section \ref{obs}). }

\begin{tabular}{c c c c c}
\hline
Instrument & Observation ID & Date (dd-mm-yyyy) & Start time (hh:mm:ss) &Exposure time (ks)  \\ \hline

IXPE &03003401&20-04-2024--23-04-2024&20:49:44& 99.8 \\
NICER &7701010101-03&20-04-2024--22-04-2024& 04:54:20&10.09 \\
VLA &24A-387&12-08-2024&05:00:42&3.53\\

\hline
\label{table1}
\end{tabular}

\end{table*}

\begin{table}
\centering

\caption{ Results obtained from the {\tt PCUBE} analysis. The uncertainties mentioned are 1 $\sigma$ error (see section \ref{model_ind_analysis}).  For the non-detections, 3 $\sigma$ upper limits are reported.}

\begin{tabular}{c c c }
\hline
&Overall&\\
Energy band& PD (\%) & PA (degree) \\ 
\hline

2-8 keV &$1.4\pm0.4$ &$-7\pm7$  \\
2-4 keV &$<$1.9&-\\
4-8 keV &$<$3.7&-\\
\hline
&Dip State&\\
Energy band& PD (\%) & PA (degree) \\ 
\hline
2-8 keV & $<$3.8 &-  \\
2-4 keV &$<$3&-\\
4-8 keV &$<$6.4&-\\
\hline
&Non-dip State&\\
Energy band& PD (\%) & PA (degree) \\ 
\hline
2-8 keV &$1.8\pm0.4$ &$1\pm6$  \\
2-4 keV &$<$2.5&-\\
4-8 keV &$2.4\pm0.7$&$3\pm8$\\
\hline

\label{table2}
\end{tabular}

\end{table}

\subsection{IXPE}
\begin{table}
\centering
\caption{Best-fitting spectral model parameters from fits to the joint NICER and IXPE spectra of GX~13$+$1 with  {\tt tbnew*(diskbb+bbodyrad+gaussian)*polconst*ionabs *ionabs*ionabs*ionabs*const} model fits to the joint NICER and IXPE spectra of GX~13$+$1. The uncertainties are 1$\sigma$. The calibration constant for NICER is fixed at unity (see Section \ref{spec_pol}).}

\begin{tabular}{c c c}
\hline
& Dip State& Non-dip state \\ 
Parameters& &  \\ 
\hline
&tbnew&\\
$\text{N}_{\rm H}$ ($10^{22}$ atoms $cm^{-2}$)&$4.18_{-0.05}^{+0.08}$&$4.21_{-0.02}^{+0.01}$ \\
$\text{N}_{\rm Si}$ ($10^{16}$ atoms $cm^{-2}$)&$2.23_{-0.10}^{+0.11}$&$2.26_{-0.06}^{+0.07}$ \\
\hline

&diskbb&\\
kT (keV)&$0.89_{-0.07}^{+0.06}$&$1.17_{-0.07}^{+0.10}$\\
Norm&$368.79_{-96.37}^{+138.31}$&$205.13_{-23.19}^{+8.50} $\\
\hline

&bbodyrad&\\
kT (keV)&$1.38_{-0.02}^{+0.02}$&$1.49_{-0.02}^{+0.02}$\\
Norm&$147.98_{-9.75}^{+7.97}$&$85.72_{-4.06}^{+9.42}$\\
\hline
&gauss&\\
$E_{\rm l}$ (keV)&$6.60^{+0.05}_{-0.06}$&$6.71^{+0.03}_{-0.03}$\\
$\sigma$ (keV)&$0.27^{+0.06}_{-0.05}$&$0.17^{+0.04}_{-0.03}$\\
Norm ($\times 10^{-3}$) &$3.06^{+0.57}_{-0.54}$&$2.98^{+0.44}_{-0.48}$\\
EW (keV)& 0.05&0.04\\
\hline
&ionabs&\\
line ID/ $E_{l}$ (keV)&2801/Ni xxviii&\\
$N_{\rm ion}$  ($\times 10^{18}$)&$6.78_{-6.15}^{+20.80}$&$2.99_{-1.85}^{+4.63}$\\
$kT_{eff}$ (keV)&Unconstrained&$8.44_{-3.87}^{+17.81}$\\
z ($\times 10^{-3}$)&$-7.17_{-2.14}^{+2.71}$&$-2.14_{-1.51}^{+1.42}$\\
\hline
&ionabs&\\
line ID/ $E_{l}$ (keV)&2802/Ni xxvii&\\
$N_{\rm ion}$  ($\times 10^{18}$)&$2.22_{-1.69}^{+4.55}$&$2.42_{-1.30}^{+1.75}$\\
$kT_{eff}$ (keV)&$3.75_{-2.82}^{+33.11}$&$6.40_{-2.44}^{+7.43}$\\
z ($\times 10^{-3}$)&$-6.51_{-1.30}^{+2.89}$&$-3.23_{-0.63}^{+0.44}$\\
\hline
&ionabs&\\
line ID/ $E_{l}$ (keV)&2602/Fe xxv&\\
$N_{ion}$  ($\times 10^{18}$)&$10.38_{-1.88}^{+2.06}$&$6.82_{-1.03}^{+1.77}$\\
$kT_{eff}$ (keV)&$4.62_{-1.61}^{+2.29}$&$17.12_{-3.83}^{+3.70}$\\
z ($\times 10^{-3}$)&$-4.28_{-0.69}^{+0.29}$&$-3.33_{-0.17}^{+0.17}$\\
\hline
&ionabs&\\
line ID/ $E_{l}$ (keV)&2601/Fe xxvi&\\
$N_{ion}$  ($\times 10^{18}$)&$20.61_{-5.69}^{+5.91}$&$15.89_{-4.71}^{+3.39}$\\
$kT_{eff}$ (keV)&$10.25_{-2.48}^{+2.84}$&$26.24_{-5.14}^{+6.52}$\\
z ($\times 10^{-3}$)&$-7.19_{-0.75}^{+0.85}$&$-4.15_{-0.30}^{+0.32}$\\
\hline
&Cross-calibration&\\
DU1&$0.925_{-0.003}^{+0.003}$&$0.907_{-0.002}^{+0.002}$\\
DU2&$0.939_{-0.003}^{+0.003}$&$0.920_{-0.002}^{+0.002}$\\
DU3&$0.908_{-0.003}^{+0.003}$&$0.892_{-0.002}^{+0.002}$\\
\hline
$\chi^{2}$/DOF&1417/1464&1463 /1484 \\
\hline

\hline
&Flux\footnote{Energy flux ($10^{-9}$ ergs/$cm^{2}$/s) at 2-8 keV energy ranges  }  &\\
 2-8 keV  &$6.36_{-0.06}^{+0.08}$ &$7.63_{-0.04}^{+0.04}$\\ 

\hline
$F_{\rm diskbb}$/$F_{\rm Total}$\footnote{Percentage of disk energy flux in the energy range 2–8 keV} (2-8)&30 &58\\
\hline

\label{table3}
\end{tabular}

\end{table}

\begin{table}
\centering
%\begin{tabular}{|c|c|c|c|c|c|p{1cm}p{1cm}p{1cm}p{1cm}p{1cm}p{1cm}p{1cm}p{1cm}p{1cm}|}
\caption{PD and PA of each spectral component obtained from the best-fit spectropolarimetric model for GX~13$+$1.  The uncertainties mentioned are at 90\% CL (see Section \ref{spec_pol}).}

\begin{tabular}{c c c}
\hline
& Dip State&  \\ 
Component& PD (\%)& PA (deg)  \\ 
\hline
bbodyrad&$4.7_{-2.2}^{+2.4}$&$-35_{-15}^{+13}$\\

diskbb&$4.2_{-3.9}^{+4.0}$&$<$29\\

\hline
bbodyrad+diskbb&$1.9_{-0.9}^{+0.9}$&$-46_{-13}^{+13}$\\
\hline
\hline
& Non-dip State&  \\ 
Component& PD (\%)& PA (deg)  \\ 
\hline
bbodyrad&$4.5_{-2.5}^{+2.6}$&$-1_{-4}^{+17}$\\
diskbb& $<$2.1 &-\\

\hline
bbodyrad+diskbb&$1.4_{-0.5}^{+0.5}$&$-1_{-9}^{+9}$\\
\hline
\label{table4}
\end{tabular}
\end{table}

IXPE observed GX~13$+$1 from 2024 April 20, 20:49:44 UTC to April 23, 01:01:27.18 UTC (PI: Unnati Kashyap) with a total on-source exposure time of approximately 99.8 ks (see Table \ref{table1} and the light curve in Figure \ref{lc}). Spectral and polarimetric analysis was performed using {\tt HEASoft version 6.33}, with the IXPE Calibration Database (CALDB) version 20240125\footnote{\url{https://heasarc.gsfc.nasa.gov/docs/ixpe/caldb/}}. Source photons were selected from circular regions with a radius of 60$\arcsec$ for I, Q, and U spectra for each Detector Unit (DU) centered at the brightest pixel located at RA of $273\fdg63$ and DEC of $-17\fdg15$. The weighted scheme NEFF was adopted for the spectro-polarimetric analysis with improved data sensitivity\footnote{\url{https://heasarc.gsfc.nasa.gov/docs/ixpe/analysis/IXPE_quickstart.pdf}} \citep{2022SoftX..1901194B, 2022AJ....163..170D}.  The ancillary response files (ARFs)
and modulation response files (MRFs) were generated for each DU
using the {\tt ixpecalcarf} task, with the same extraction
radius used for the source region. Since GX~13$+$1 is a bright source, no background rejection and subtraction schemes were implemented,
following the prescription by \cite{2023AJ....165..143D}. The unweighted model-independent polarimetric analysis was performed using the {\tt IXPEOBSSIM package version 31.0.1} \citep{2022SoftX..1901194B}. For extracting images and spectra, {\tt XSELECT} available as a part of the {\tt HEASoft 6.33} package was used.

\subsection{NICER}
The Neutron Star Interior Composition Explorer (NICER) observed GX 13+1 from 2024 April 20, 04:54:20 UTC to 2024 April 23, 23:33:40 UTC. The observation details are summarized in Table \ref{table1}. The NICER/XTI observations were reduced using the NICER software {\tt NICERDAS} distributed with {\tt HEASoft 6.33}, CALDB 20240206\footnote{\url{https://heasarc.gsfc.nasa.gov/docs/heasarc/caldb/nicer/}}, and updated geomagnetic data. Cleaned event files were generated using the {\tt nicerl2} pipeline and by applying standard filtering criteria, limiting undershoot rates to $<$ 500 cts/s and overshoot rates to $<$ 30 cts/s. The {\tt nicerl3-spect} task was employed to generate source spectra and background spectra using the {\tt SCORPEON} background model, along with the detector responses. We note that the observations were carried out during orbit day and orbit night observations. Since the orbit day data are often affected by the optical light leak, we only used orbit night observations for the spectro-polarimetric analysis.

\subsection{The Very Large Array}

GX~13$+$1 was observed with the Karl G. Jansky Very Large Array (VLA) under Project Code 24A-387 on August 12 2024, with $\sim$1 hour on source at X-band (3-bit samplers at 8--12\,GHz with 4\,GHz of bandwidth). The data were obtained from the National Radio Astronomy Observatory (NRAO) data archive with a standard VLA pipeline calibration applied. The flux, bandpass, and polarization angle calibrator was 3C286 (J1331+030), the complex gain calibrator was J1832-1035, and the polarization leakage calibrator was J1407+2827. Data were processed in Common Astronomy Software Applications (CASA; \citep{2022PASP..134k4501C}) and calibrated for polarization using standard tasks. Data were imaged with \texttt{tclean} and phase self-calibrated.

\section{Results}
\subsection{Source Spectral State}

\begin{figure}
\centering
\includegraphics[width=0.45\textwidth]{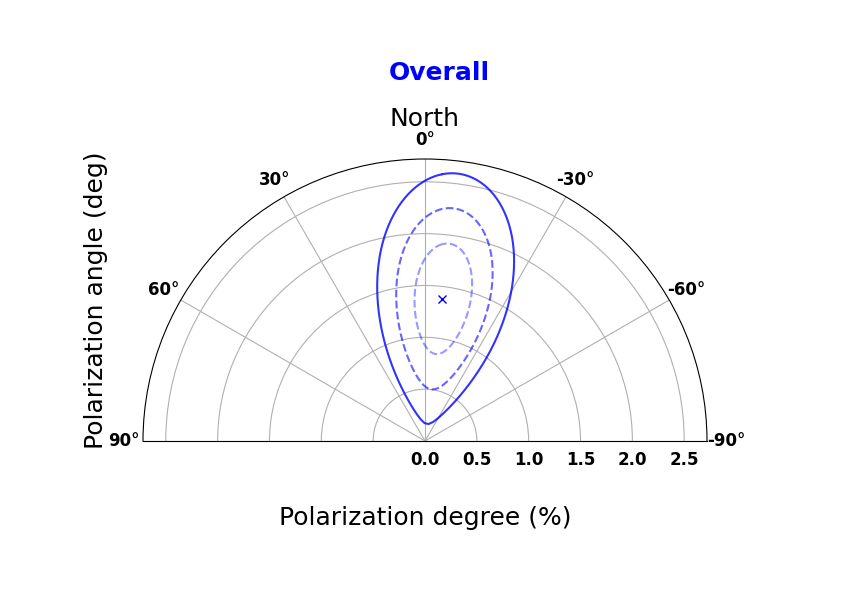}
\includegraphics[width=0.45\textwidth]{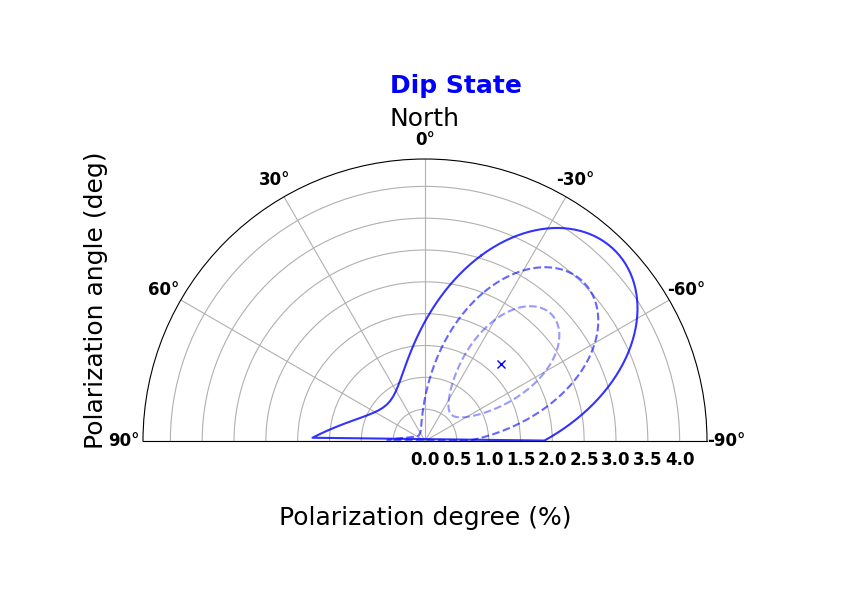}
\includegraphics[width=0.45\textwidth]{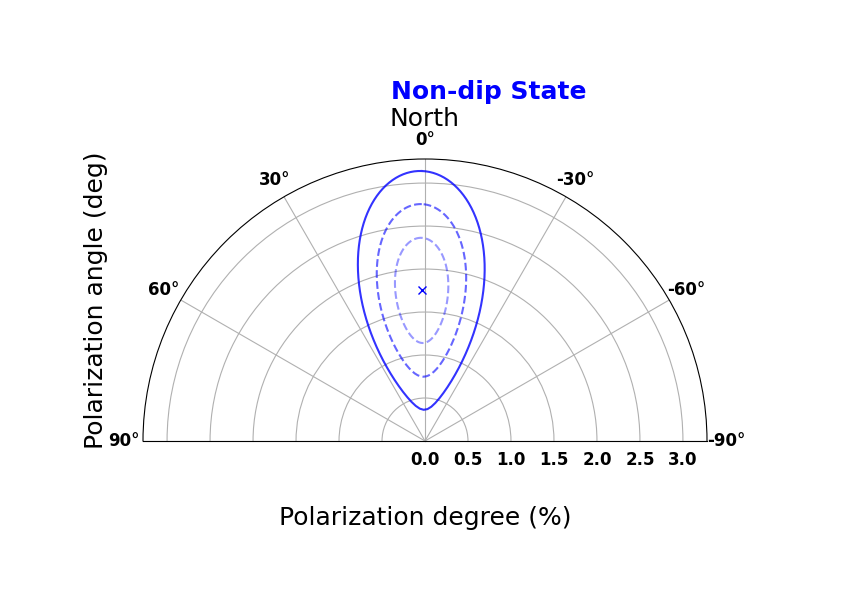}

\caption{Contour plots of the polarization degree and angle, determined with the {\tt PCUBE} algorithm, at the 68 \%, 95 \% and 99 \% confidence levels, in the 2–8 keV energy band during the overall (upper panel), dip state (middle panel) and, non-dip state (lower panel).}
\label{polar_mod_ind}
\end{figure}

\begin{figure}
\centering
\includegraphics[width=0.5\textwidth]{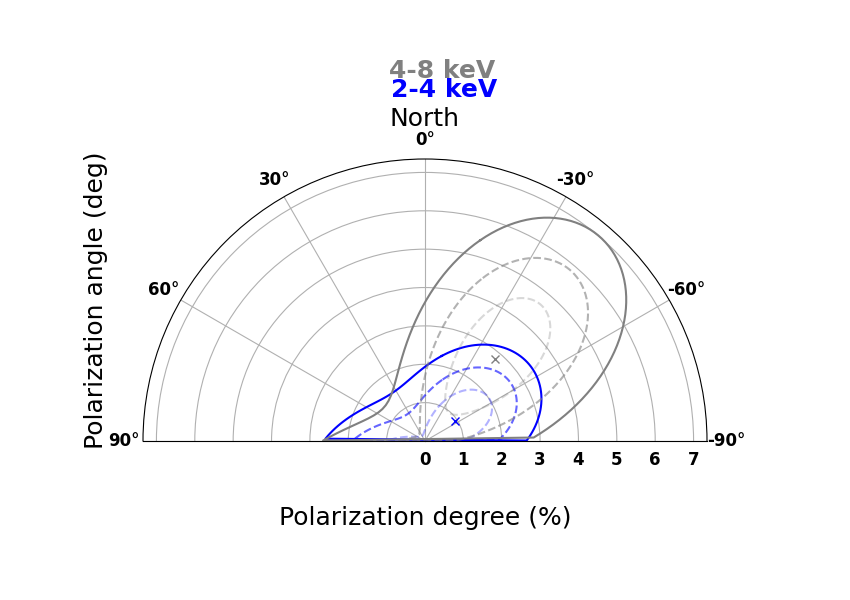}
\includegraphics[width=0.5\textwidth]{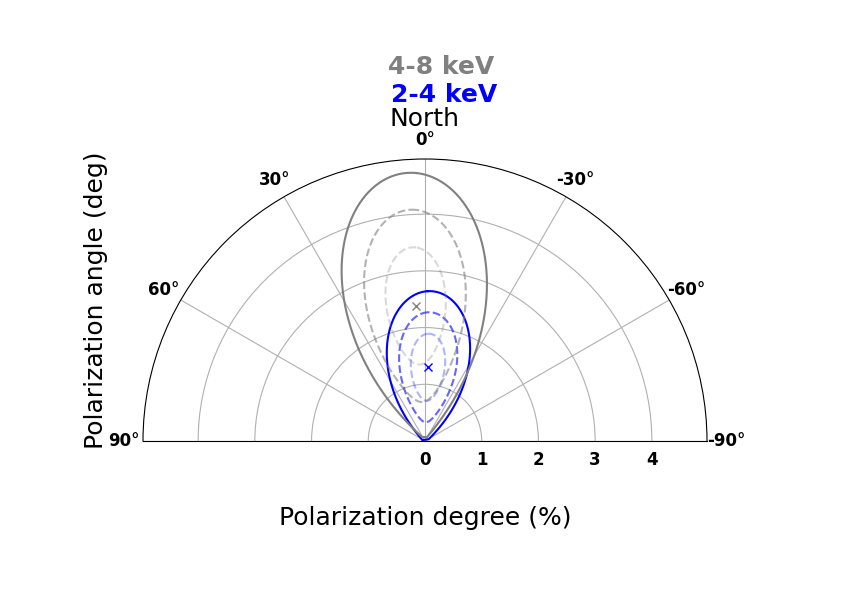}
\caption{Contour plots of the polarization degree and angle, determined with the {\tt PCUBE} algorithm, at the 68 \%, 95 \% and 99 \% confidence levels, in the 2–4 keV and 4-8 keV energy band during the dip state (upper panel) and non-dip state (lower panel).}
\label{polar_mod_ind_en}
\end{figure}

\begin{figure*}
\centering
\includegraphics[width=0.45\textwidth]{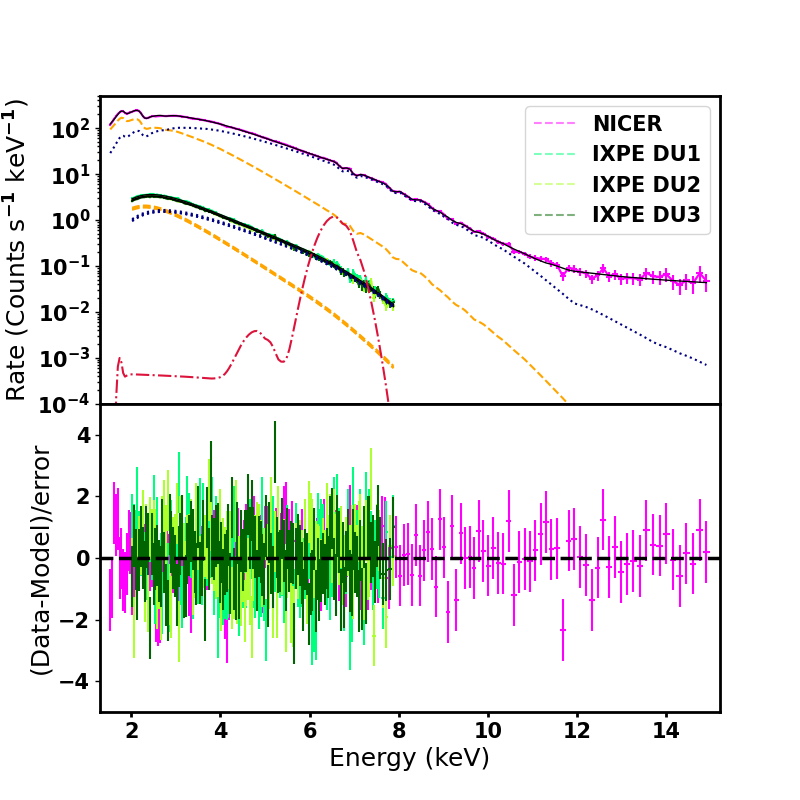}
\includegraphics[width=0.45\textwidth]{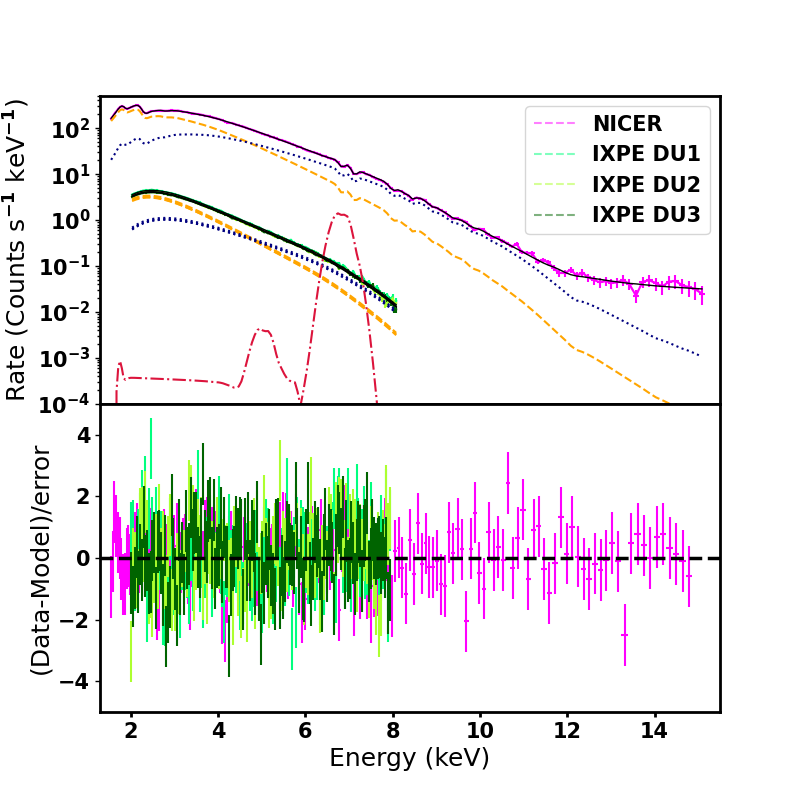}

\caption{Model fitted joint spectra of GX~13$+$1 as observed by IXPE DU1 (spring green), IXPE DU2 (green yellow), IXPE DU3 (dark green), and NICER (magenta) during the dip state (left) and non-dip state (right). The spectra are fitted with the {\tt tbnew*(diskbb+bbodyrad+gauss)*polconst*ionabs*ionabs*ionabs*ionabs*const} model in the 2-10 keV energy band. The total model is shown with the solid black (NICER and IXPE), and the individual additive components {\tt diskbb}, {\tt bbodyrad}, and {\tt gauss} are shown with the dashed orange, dotted navy, and dashdot crimson lines, respectively. The lower subpanel shows the residuals between the data and the best fit model.}
\label{spec_all}
\end{figure*}

Figure \ref{lc} shows the IXPE and NICER light curves of the source GX~13$+$1, which exhibits multiple dips in the light curve. Figure \ref{hid} shows the NICER Hardness Intensity Diagram (HID), where the hardness is defined as the ratio of the count rates in two energy bands 3.8-6.8 keV and 2.0-3.8 keV. Although the source spectral state is not discernible from the NICER HID, the shape of the HID suggests the source was in parts of the Z-track during the IXPE observations.  A comparison of the NICER HID with the previously reported study of the source using the same NICER observations (see Figure 1 of \cite{2024RNAAS...8..243K}) hints towards the source being primarily in the FB of the Z-track during the IXPE observations.  
The HID shows a trend as the source moves from the low-intensity to the high-intensity state in the light curve.
The dashed line in Figure \ref{lc} shows the transition between the two intensity (low and high) levels, hereafter dip and non-dip state, which is also consistent with the dip and non-dip state shown in Figure \ref{hid}.  We define the dip state as periods when the source count rate falls below 7.5 counts/s in the IXPE light curve and remains at this level for an extended period of time, lasting several hours. We note here that we include the adjacent time intervals during which the count rate transitions into or out of this state (i.e., ingress and egress phases) as a dip state. All remaining segments of the light curve are classified as the non-dip state. The total IXPE exposure that falls within the dip and the non-dip states are 26.9 ks and 72.9 ks.
To investigate the source geometry during the dip and non-dip states, we carry out detailed spectro-polarimetric investigations for each of these states separately.

\subsection{Model-independent polarimetric analysis}
\label{model_ind_analysis}

We performed a polarimetric analysis of GX~13$+$1 using the {\tt ixpeobssim} package \citep{2022SoftX..1901194B} under the {\tt PCUBE} algorithm in the {\tt xpbin} tool \citep{2015APh....68...45K}. The unweighted analysis on the data implemented in the {\tt ixpeobssim} package was applied. We employed the polarimetric analysis of the data in the 2–8 keV, 2-4 keV, 4-6 keV, 4-8 keV, and 6-8 keV energy bands, and the results obtained are reported in Table \ref{table2}. The results obtained from the model-independent analysis show significant detection of polarization in the 2-8 keV energy band during both dip and non-dip states (see Table~\ref{table2}).

\subsection{Spectro-polarimetric analysis}
\label{spec_pol}
We carried out spectral fitting and statistical analysis using {\tt XSPEC v 12.14.0h} spectral fitting package distributed as a part of the {\tt Heasoft 6.33 package}.  Considering the IXPE and NICER observations for the dip and non-dip spectra, we fit the joint IXPE and NICER spectra with a model consisting of a multicolour disk blackbody model (MCD; {\tt diskbb} in XSPEC) and a blackbody radiation model ({\tt bbodyrad} in {\tt XSPEC}) in the 1.5–15.0 keV energy range. The total spectrum was modified by the presence of neutral hydrogen absorption in the interstellar medium, and this was accounted for by using the {\tt tbnew}\footnote{\url{https://pulsar.sternwarte.uni-erlangen.de/wilms/research/tbabs/}} model. The spectral fits show a significant residual around $\sim$ 1.8~keV, which is likely the Si K edge, a known NICER instrumental systematic\footnote{\url{https://heasarc.gsfc.nasa.gov/docs/nicer/data_analysis/workshops/2024/joint2024.html}}. The Si abundance, when allowed to vary relative to the \texttt{wilms} abundance, was found to have a value of $\sim$2.2 $\times 10^{16}$ atoms ${\rm\,cm}^{-2}$, which improved the fits significantly. The abundances and photoelectric cross-sections are adopted from \citep{2000ApJ...542..914W}. The iron (emission) features detected in the spectrum of both the dip and non-dip states are described by an additional Gaussian ({\tt gauss}) component.  A constant ({\tt const}) model was used to account for the uncertainties in cross-calibration uncertainties between NICER and the IXPE DUs and is reported in Table \ref{table3}. 

  The previous studies shows that the moderate-resolution spectra of GX~13$+$1 from ASCA exhibited highly ionized iron K absorption lines \citep{2001ApJ...556L..87U}. Further high-resolution spectroscopy with Chandra/HETGS observations showed the presence of K$\alpha$ lines from H-like, Fe, Mn, Cr, Ca, Ar, S, Si, and Mg as well as He-like Fe. The blue-shifted lines showed outflowing materials with velocity $v_{out} \sim$ 400 $\rm {km/s}$ \citep{2004ApJ...609..325U,2014MNRAS.438..145M,2018ApJ...861...26A,2020MNRAS.497.4970T}. 

Our analysis shows that the NICER+IXPE joint spectra of GX~13$+$1 during the dip and the non-dip state exhibit several absorption features. We modified the fit using four {\tt ionabs} (XRISM collaboration, in prep) models describing these features. We note here that {\tt ionabs}  is a table model, which is the extended version of {\tt Kabs} \citep{1999JQSRT..62...29W,2020MNRAS.497.4970T}. It accounts for the absorption lines and the edges from a single ion that includes the higher-order transition from ground level.  The lines in the {\tt ionabs} model are calculated considering a Voigt profile, using a convolution of the Gaussian and the Lorentzian function. The  {\tt ionabs} model parameters include an ID of ion, ion column density $N_{\rm ion}$ in units of $10^{18} cm^{-2}$, an effective temperature for the Doppler broadening $kT_{\rm eff}$ in units of keV, and redshift z. The ion ID is written as (atomic number)+(the number of electrons). The effective temperature $kT_{\rm eff}$ is written as the $2\times$$kT_{\rm eff}$ /$m_{\rm atom}$ = 2kT/$m_{\rm atom}$ +$v_{\rm turb}^{2}$, where T is the real temperature of the gas and $v_{\rm turb}$ is the microscopic turbulent velocity
of the gas.

To study the polarization of the spectral components, first, we applied the {\tt polconst} model to the entire continuum model to check the consistency with the model-independent analysis. The PD and the PA obtained are consistent with the upper limits of PD and PA obtained from the model-independent analysis (See section \ref{model_ind_analysis}). The results obtained from the model-independent studies show evidence of possible PA swings between the dip and the non-dip state and a hint of increasing PD with energy has been observed in the non-dip state. Thus, to further disentangle the relative contribution to the polarization signal from the two spectral components (soft and hard)  to the total PD and PA observed during the dip and the non-dip state, we applied a model with different polarizations for each spectral component, and the obtained polarization for the {\tt diskbb}, and the {\tt bbodyrad} component at 90\% confidence level are reported in Table \ref{table4}. Figure~\ref{spec_all} shows the NICER (Magenta) and IXPE (Spring green, Green yellow, and Dark green) spectra along with the best-fitting models during the dip (upper panel) and non-dip (lower panel) states, and the corresponding best-fitting values are reported in Table \ref{table3} and \ref{table4}. We note here that the results obtained from the spectro-polarimetric analysis are consistent with the results reported in \cite{2025ApJ...979L..47D}, taking the uncertainties into account. The observed discrepancy in the reported disk blackbody normalization values may be due to the different absorption models employed in our analysis. We also note that the upper limit (at the 90\%  CL) obtained for the {\tt diskbb} PA ($<$ 29) for the dip state and PD ($\sim$ 4.2) obtained for the dip state obtained in our case are found to be slightly lower than the upper limits reported in  \cite{2025ApJ...979L..47D}.

\begin{figure}
\centering
\includegraphics[width=0.4\textwidth]{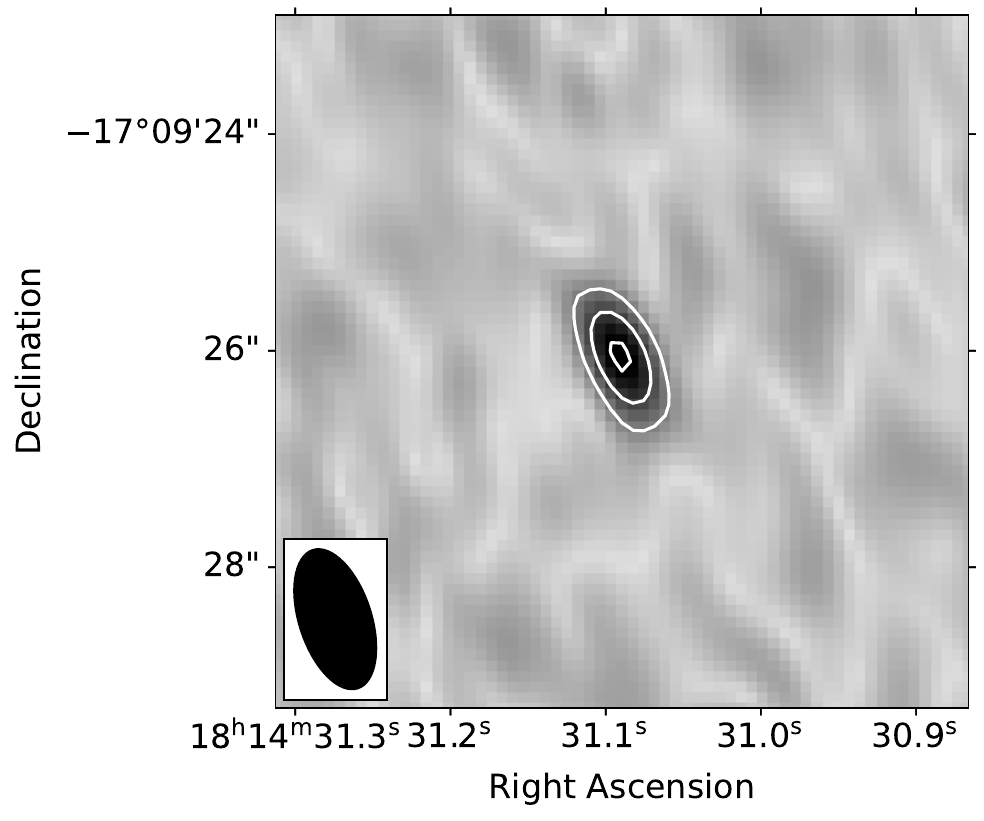}
\caption{Image of the linear radio polarization detection of GX~13+1 at a frequency of 10\,GHz. Contours plotted are equal to 3, 5, and 7$\sigma$ flux density levels, where $\sigma$ is the background RMS value of $5.6~\mu$Jy/beam. The synthesized beam is shown in the lower left corner.}
\label{fig:radio_lpoli}
\end{figure}

\section{Radio Observations} 

GX~13$+$1 behaves like a Z source in radio with variable but persistent radio emission, which we attribute to jet-like radio emission from canonical Z sources. 

GX~13+1 was detected in radio with the VLA at 10~GHz with a peak Stokes~I flux density of $660.3 \pm 11.9$~$\mu$Jy and with a total linear polarization flux density of $40.8 \pm 5.6$~$\mu$Jy (Figure~\ref{fig:radio_lpoli}), for a polarization degree of $6.2 \pm 0.9$\%. The observed polarization angle (prior to Faraday de-rotation) was $-59^{\circ} \pm 3^{\circ}$. To account for the Faraday rotation of the radio emission (frequency-dependent rotation of the polarization angle as the radio wave travels through a parallel magnetic field), we split the $8-12$\,GHz observed frequency band into two subbands of 2\,GHz each ($8-10$ and $10-12$\,GHz bands). We estimate the intrinsic polarization angle of the radio emission, $\Phi_{\mathrm{0}} = \Phi_{\mathrm{obs}} + \mathrm{RM}\,\lambda^{2}$, where $\mathrm{RM}$ is the rotation measure and $\lambda$ is the photon wavelength. We obtain a rotation measure of $\sim163~\mathrm{rad~m^{-2}}$, and thus an intrinsic, Faraday de-rotated jet position angle of $-68^\circ \pm 3^\circ$. We also note that there is another observation epoch of GX~13$+$1 where radio polarization is detected, which is lower in signal:noise but is consistent in the polarization angle and rotation measure to the data presented here, and further indicates that the intrinsic jet position angle is constant in time and not precessing, as we already assumed. Further discussion and analysis of the radio polarization in GX~13+1 will be presented in Pattie et al. (in prep).

\section{Discussion}
\label{discussion}
\begin{table}
\centering
%\begin{tabular}{|c|c|c|c|c|c|p{1cm}p{1cm}p{1cm}p{1cm}p{1cm}p{1cm}p{1cm}p{1cm}p{1cm}|}
\caption{PD and PA of {\tt bbodyrad} component obtained from the best-fit spectropolarimetric model for GX~13$+$1, assuming a source inclination of 70\degr and a geometry with the disk plane perpendicular to the radio jet position angle. The uncertainties mentioned are at 90\% CL (see Section \ref{discussion}).}

\begin{tabular}{c c c}
\hline
& Dip State&  \\ 
Component& PD (\%)& PA (deg)  \\ 
\hline
bbodyrad&$4.2_{-1.3}^{+1.3}$&$-51_{-9}^{+9}$\\
diskbb& 3.0&22\\
\hline
\hline
& Non-dip State&  \\ 
Component& PD (\%)& PA (deg)  \\ 
\hline
bbodyrad&$3.2_{-1.2}^{+1.2}$&$-37_{-11}^{+11}$\\
diskbb& 3.0&22\\
\hline
\label{table5}
\end{tabular}
%\caption{Observation Details (SXT)}
%\label{table2}
\end{table}

\begin{figure}
\centering
\includegraphics[width=0.45\textwidth]{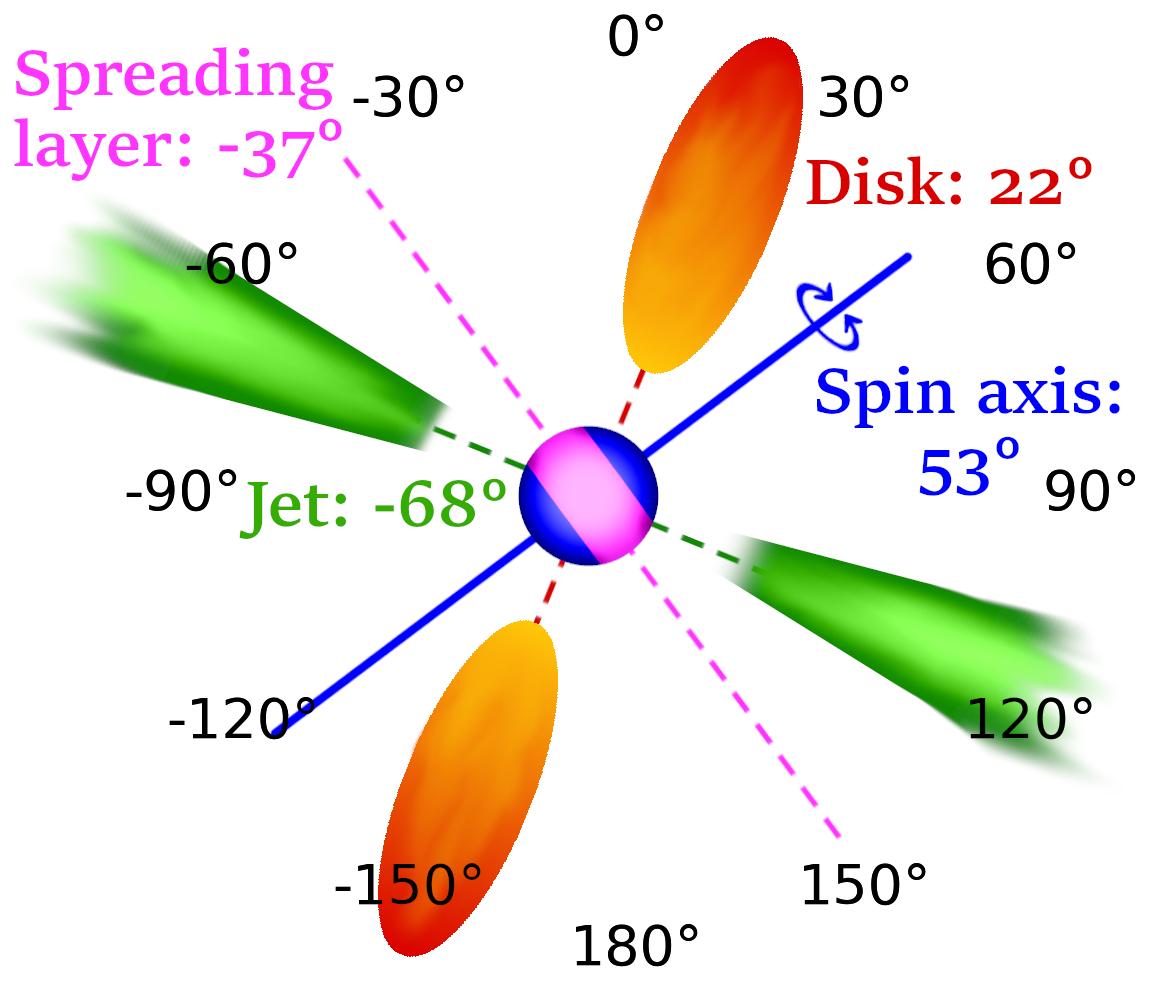}

\caption{Schematic representation of the geometry of GX~13$+$1 based on our X-ray and radio polarimetry. The radio-jet with a position angle of $\sim$ -68\degr is shown in green. The disk (PA=$22\degr$) assumed to be oriented orthogonally with respect to the radio jet is shown in red. The BL/SL with a PA of $\sim$ -37\degr representing the Comptonization region is shown in Magenta. Assuming the BL/SL spreading coplaner to the accretion disk, we find a misalignment between the estimated NS spin and the orbital axis of GX~13$+$1.} 
\label{schematic}
\end{figure}

In this paper, we report the X-ray and the first radio polarization studies of the NS source GX~13$+$1 using IXPE and VLA. As shown in Figure~\ref{lc}, the source exhibits clear flux variability, along with the presence of multiple dips in the light curve during our IXPE observations, consistent with the previously reported behavior of the source \citep{2024A&A...688A.217B,2024A&A...688A.170B}. A simultaneous NICER observation shows a similar pattern in the light curves, exhibiting a trend in the HID distinguishing the dip and non-dip states (see Figure~\ref{hid}). The NICER HID indicates the source tracing out parts of Z-track during our IXPE observations, consistent with the previously reported NICER studies of the source \citep{2018ApJ...861...26A,2023MNRAS.522.3367S,2024RNAAS...8..243K}. 

The source spectra during the dip and non-dip state are well described by a {\tt diskbb} component, representing the softer emission from an accretion disk, along with a {\tt bbodyrad} component, representing the hard Comptonized emission attributed to a boundary layer (BL) between the disk and the NS or a vertically extended spreading layer (SL) around the NS. The source spectra during our observation is observed to be consistent with the previously reported spectral shape of the source \citep{2024A&A...688A.217B,2024A&A...688A.170B}. We note here that the addition of a Gaussian ({\tt gauss}) component representing the emission feature detected, along with four {\tt ionabs} components representing Fe xxv, Fe xxvi, Ni xxvii, and Ni xxviii absorption lines representing the absorption in the wind above the disk provides a relatively better-fit statistics \citep{2018ApJ...861...26A,2020MNRAS.497.4970T}. The spectral studies show that the contribution of the disk was relatively low during the dip state (30\%) in comparison to the non-dip state, where the disk contributes to almost 58\% of the total flux observed from the source. The presence of the dips during the states where the Comptonization component dominates may suggest a possible association between this component and the observed dips from the source.

The model-independent studies of GX~13$+$1 show a polarization of ${\rm PD} = 1.7 \pm 0.7\%$ with a polarization angle of ${\rm PA} = -44 \pm12 \degr$ during the dip and ${\rm PD} = 1.8 \pm 0.4\%$ with a polarization angle of ${\rm PA} = 1 \pm6 \degr$ during the non-dip state. The X-ray spectro-polarimetric analysis shows polarization of ${\rm PD} = 1.9 \pm 0.9\%$ with a polarization angle of ${\rm PA} = -46 \pm13 \degr$ during the dip state, and a polarization of ${\rm PD} = 1.4 \pm 0.5\%$ with a polarization angle of ${\rm PA} = -1 \pm9 \degr$ during the non-dip state, consistent with the results obtained from the model-independent polarimetric studies. Figure~\ref{polar_mod_ind} and Table~\ref{table2} clearly show that the source exhibits a PA swing between the dip and non-dip states, along with a hint of increasing PD with energy during the non-dip state (see  Table~\ref{table2} and Figure~\ref{polar_mod_ind_en}). However, in the dip state, no energy-dependent variation of PD has been detected. \cite{2024A&A...688A.170B} reported a similar trend of PD rising with energy during the higher flux state and a constant PD during the lower flux state during the previous observations of GX~13$+$1. Our observations also show a high disk PD, when the contribution of the disk to the total spectrum is low, consistent with the PD variations reported in \citep{2024A&A...688A.217B}.

IXPE has studied several NS LMXBs, and among them, GX~13$+$1 is known to be the peculiar one exhibiting dramatic variations of polarization properties with strong PA rotations. The previous observations of the source report a continuous PA rotation \citep{2024A&A...688A.170B}.  However, a more recent study reports a rather complex polarization behavior correlated with the hardness ratio and source intensity \citep{2025ApJ...979L..47D}. A comparative study of all three IXPE observations shows a similar evolution of the polarimetric properties across all three observations with a similar PA among the pre-dip and post-dip intervals  (see Figure 7 of \cite{2025ApJ...979L..47D}). Previous studies report scattering in the ADC or scattering in the disk wind is reported to be associated with the observed dips, driving the switching polarization properties observed from the source \citep{2024A&A...688A.170B,2025ApJ...979L..47D}. A similar argument has been made for the BH sources exhibiting higher polarization at relatively higher energy band \citep{2023ApJ...949L..43R}(see also discussions of \cite{2024ApJ...964...77R}). However, we note here that we are limited by the sensitivity of the data presented in this work, needed for further investigating and distinguishing polarizations across different energy bands (see Table~\ref{table2}).

Here we propose an alternative possibility, where the spectral component presumably associated with the scattering likely remains invariant between the dip and the non-dip state. In contrast, the contribution from the emission components associated with the softer disk and the harder BL/SL components vary as the source transitions from the dip to the non-dip state. During the dip state, the source flux diminishes, either due to stochastic obscuration or due to the geometrical evolution of the disk. The change in the disk contributions observed in our analysis (see Table~\ref{table2}) during the dip and the non-dip state also supports this possibility. Consequently, the switching polarization properties during the dip and the non-dip state are possibly associated with the changing relative contributions of the emission components. Hence, our current understanding of the source manifests the need for more realistic modeling that includes all three components: the disk, the Comptonization region, and the scattering medium.

\subsection{The possible source geometry based on the X-ray and radio polarimetry}

We further discuss the possible accretion geometry of the source in light of our X-ray and the radio polarization results. Our radio observations with the VLA show a variable but persistent jet emission from GX~13$+$1. While the jet is variable in flux density, the jet position axis is expected to remain constant over time as there is no evidence of precession of the inner or outer accretion disk, nor of the jet, and the potential effect of QPOs is too small to affect the average jet axis given its physically large scale \citep{Fomalont2001ApJ...558..283F}.

We measured a Faraday-corrected radio polarization angle of $-68\degr \pm 3\degr$ in GX~13$+$1. The timing properties indicate that the jet is a compact, conical jet similar to typical hard state jets. In this case, the PA can be assumed to be aligned with the jet axis \citep[e.g.,][]{Corbel2000A&A...359..251C, Russell2015MNRAS.450.1745R}. However, the spectral index of the radio emission is steep at $\sim-0.65$, in which case the observed PA is assumed to be perpendicular to the magnetic field, and thus perpendicular to the jet axis \citep[e.g.,][]{Fender1999MNRAS.304..865F, Curran2014MNRAS.437.3265C, Hughes2023MNRAS.521..185H}. Due to the highly misaligned nature of the X-ray and radio PA (close to $45^\circ$), whether the radio PA is aligned or perpendicular to the jet axis does not make a significant difference in the conclusions of this work. A more detailed analysis of the radio properties of this source will be presented in a future work (Pattie et al. in prep.). We therefore take the aligned PA and jet axis as a test case, and note that a perpendicular radio jet is also a possibility, with similar conclusions.

Assuming that this radio polarization angle is parallel to the jet axis,  
we investigate the X-ray polarization properties of the soft and hard emission components during the dip and non-dip states of GX~13$+$1. 

The plane of the accretion disk is expected to be oriented orthogonally with respect to the position angle of the radio jet \citep{1982MNRAS.199..883B}. Therefore, assuming a disk with a ${\rm PA} = 22 \degr$ and a ${\rm PD} = 3.0$ obtained for a standard
electron-scattering dominated optically thick accretion
disk \citep{1960ratr.book.....C}, for a source inclination of 70\degr \citep{2014A&A...564A..62D,2023MNRAS.522.3367S}, we estimate the polarizations associated with the Comptonization ({\tt bbodyrad})  component during both the dip and the non-dip states. We report the polarization results obtained from the spectro-polarimetric analysis under this assumption in Table \ref{table5}. 

With our assumptions, we find that the Comptonization component is polarized with ${\rm PD} = 3.2 \pm 1.2\%$ at a polarization angle of ${\rm PA} = -37 \pm11 \degr$ during the non-dip state and with  ${\rm PD} = 4.2 \pm 1.3\%$ at a polarization angle of ${\rm PA} = -51 \pm9 \degr$ during the dip state. Thus, the estimated PD attributed to the Comptonized component during both the dip and the non-dip states are observed to decrease slightly than the values reported in Table \ref{table4} and are consistent with the polarization associated with a region of moderate optical depth \citep{1985A&A...143..374S}. Moreover, the polarization attributed to the Comptonized component is observed to be higher during the dip state relative to the non-dip state.

 Numerical simulations predict the PD associated with the SL-like Comptonization region in the case of WMNS as $<$ 1.5\% \citep{2025A&A...696A.181B} and the PD associated with the BL-like Comptonization region as $<$ 0.5\% \citep{2024A&A...684A..62F}. In the case of GX~13$+$1, the observed PD (for both the dip and the non-dip state) associated with the Comptonization component is relatively higher than the PD predicted by the numerical simulations and cannot be explained by repeated Compton scattering in high optical depth environments \citep{2025A&A...696A.181B}. As reported in previous IXPE studies, a complementary possibility could be the reflection of the Comptonized photons from the disk atmosphere or scattering in the wind above the accretion disk as a possible contributor resulting in a relatively higher polarization, with the SL at the NS surface serving as a source of photons \citep{2024Galax..12...43U}. Another possibility could be a Comptonized component with a slab-like corona, along with a relatively soft disk emission presented in \citep{2022MNRAS.514.2561G}, with an unpolarized BL component.

 During the dips, the polarization or geometry is expected to be more complex, with no known physical mechanisms associated with it. Hence, we further examine the non-dip state polarization results only, as they may provide a more accurate representation of the possible source geometry.
As reported in  Table \ref{table5}, during the non-dip state, the hard Comptonized component ({\tt bbodyrad}) PA is observed to be oriented at an angle of $\sim$ 50\degr with respect to the soft disk component ({\tt diskbb}). Typically, in the case of aligned systems, we expect the polarization vectors representing the emissions from the disk and the Comptonized component to be either nearly parallel or perpendicular to each other. Thus, a $\sim$ 50\degr (see Table \ref{table5}) PA difference obtained between the disk and the Comptonized component provides hints of possible misalignments in the case of GX~13$+$1. Furthermore, under these assumptions, considering the BL parallel to the NS equator, the spin axis of the NS in the case of GX~13+1 is found to be essentially misaligned ($\sim$ 59 \degr) with respect to the orbital axis. Figure \ref{schematic} demonstrates the possible geometry of GX~13$+$1 with the spin-orbit misalignment. Thus, our X-ray and radio polarimetric study provides evidence of spin-orbit misalignment in the case of NS LMXB GX~13$+$1. 

The spin-orbit misalignment in X-ray binary sources has been theoretically predicted previously and is suggested to be caused by the high velocity
they acquire during their formations \citep{2010ApJ...719L..79F,2019MNRAS.489.3116A}. Additionally, previous timing studies have reported detections of QPOs and Horizontal Branch Oscillations from GX~13$+$1 \citep{1998ApJ...499L..41H,2003A&A...406..221S,2024MNRAS.52711855G} and such detections are often associated with the presence of the Lense-Thirring Precession effect \citep{1998ApJ...492L..59S}, thought to have originated due to the misalignment of the accretion disk to the spin of the central object. Perhaps the detection of QPOs provides independent evidence of the presence of Lense-Thirring Precession, and hence the spin-orbit misalignment in GX~13$+$1.

\subsection{Misalignment of the disk axis with the NS axis: A comparison with the NS Z-sources}

 Considering the absence of strong GR effects and different geometries of the scattered BL radiation and the intrinsically polarized disk radiation (e.g., \cite{1960ratr.book.....C}), the disk component is often assumed to be orthogonal to the Comptonization component. The polarization properties of the individual emission (soft and hard) components have been constrained under this assumption for multiple Z-type NS LMXBs such as  Sco X-1 \citep{2024ApJ...960L..11L}, XTE J1701$-$462 \citep{2023A&A...674L..10C}, Cyg X-2 \citep{2023MNRAS.519.3681F}, and GX 349$+$2 \citep{2025ApJ...986..207K}, where spectro-polarimetry could not determine the polarization angles (PAs) of the emission components fully independently due to limited data sensitivity (see also \cite{2025A&A...699A.230G} for a recent review).  However, a difference in the PA attributed to the soft and hard emitting components has been reported for a few other Z-type LMXBs, where the data sensitivity was sufficient to estimate the PAs of the individual components.  For example, in the case of GX 5-1, the spectropolarimetric analysis shows that the disk and the Comptonization components have nonorthogonal PAs, likely resulting in an unexpected energy-dependent PA variation observed from the source \citep{2024A&A...684A.137F}. The spectropolarimetric analysis reported in \cite{2024A&A...691A.253L} shows an uncommon PA difference of $\sim$40\degr between the disk and the Comptonization components in the case of GX~340$+$0 observed in HB. However, this could not be investigated further in the NB state observation of the source, due to the limited sentivity \citep{2024arXiv241100350B}. The reported differences in PA observed during the HB state of GX~340$+$0 are similar to the differences reported for Cir~X-1 \citep{2024ApJ...961L...8R} and previous study of GX~13+1 \citep{2024A&A...688A.217B}. Previous studies also suggest that obtaining a difference in the PA between the soft and the hard components-- as we report here for GX~13+1--requires some degree of misalignment of the disk axis with the NS axis. This is also consistent with the results we obtained from the comparison of the radio and X-ray polarization findings in this study.

\section{Summary}
 In this work, we report the X-ray and first radio polarization study of the peculiar NS LMXB GX~13$+$1 using IXPE and the VLA. The X-ray polarimetric results show a source geometry comprising of a soft disk component, along with a hard Comptonization component associated with a BL/SL, consistent with the previously reported spectral shape of the source. Our VLA radio observations show variable but persistent radio jet emissions with a total linear polarization flux density of $40.8 \pm 5.6$~$\mu$Jy for a polarization degree of $6.2 \pm 0.9$\%. The X-ray and radio polarimetric data presented in this work, suggest a complex geometry of GX~13$+$1 with misaligned spin and orbital axis.

\section{Acknowledgments}
\begin{acknowledgments}
This research used data provided by the Imaging X-ray Polarimetry Explorer (IXPE), NICER (Neutron star Interior Composition Explorer), and VLA (Very Large Array) and distributed with additional software tools by the High-Energy Astrophysics Science Archive Research Center (HEASARC), at NASA Goddard Space Flight Center (GSFC). U.K. and T.J.M.  acknowledges support by NASA grant 80NSSC24K1747. M.N. is a Fonds de Recherche du Quebec – Nature et Technologies (FRQNT) postdoctoral fellow. We acknowledge Ryota Tomaru and Chris Done for sharing the {\tt ionabs} model packages with us, which helped us significantly in the spectro-polarimetric analysis.

\end{acknowledgments}

\vspace{5mm}
\facilities{IXPE, NICER, VLA}


\software{ixpeobssim (Baldini et al. 2022), xspec
(Arnaud 1996), HEASoft (Blackburn 1995), CASA (Team et al. 2022)
          }



%\appendix


\bibliography{sample631}{}

\begin{thebibliography}{}
\expandafter\ifx\csname natexlab\endcsname\relax\def\natexlab#1{#1}\fi
\providecommand{\url}[1]{\href{#1}{#1}}
\providecommand{\dodoi}[1]{doi:~\href{http://doi.org/#1}{\nolinkurl{#1}}}
\providecommand{\doeprint}[1]{\href{http://ascl.net/#1}{\nolinkurl{http://ascl.net/#1}}}
\providecommand{\doarXiv}[1]{\href{https://arxiv.org/abs/#1}{\nolinkurl{https://arxiv.org/abs/#1}}}

\bibitem[{{Allen} {et~al.}(2018){Allen}, {Schulz}, {Homan}, {Neilsen}, {Nowak}, \& {Chakrabarty}}]{2018ApJ...861...26A}
{Allen}, J.~L., {Schulz}, N.~S., {Homan}, J., {et~al.} 2018, \apj, 861, 26, \dodoi{10.3847/1538-4357/aac2d1}

\bibitem[{{Atri} {et~al.}(2019){Atri}, {Miller-Jones}, {Bahramian}, {Plotkin}, {Jonker}, {Nelemans}, {Maccarone}, {Sivakoff}, {Deller}, {Chaty}, {Torres}, {Horiuchi}, {McCallum}, {Natusch}, {Phillips}, {Stevens}, \& {Weston}}]{2019MNRAS.489.3116A}
{Atri}, P., {Miller-Jones}, J.~C.~A., {Bahramian}, A., {et~al.} 2019, \mnras, 489, 3116, \dodoi{10.1093/mnras/stz2335}

\bibitem[{{Baldini} {et~al.}(2022){Baldini}, {Bucciantini}, {Lalla}, {Ehlert}, {Manfreda}, {Negro}, {Omodei}, {Pesce-Rollins}, {Sgr{\`o}}, \& {Silvestri}}]{2022SoftX..1901194B}
{Baldini}, L., {Bucciantini}, N., {Lalla}, N.~D., {et~al.} 2022, SoftwareX, 19, 101194, \dodoi{10.1016/j.softx.2022.101194}

\bibitem[{{Bandyopadhyay} {et~al.}(2002){Bandyopadhyay}, {Charles}, {Shahbaz}, \& {Wagner}}]{2002ApJ...570..793B}
{Bandyopadhyay}, R.~M., {Charles}, P.~A., {Shahbaz}, T., \& {Wagner}, R.~M. 2002, \apj, 570, 793, \dodoi{10.1086/339776}

\bibitem[{{Barret}(2001)}]{2001AdSpR..28..307B}
{Barret}, D. 2001, Advances in Space Research, 28, 307, \dodoi{10.1016/S0273-1177(01)00414-8}

\bibitem[{{Bhargava} {et~al.}(2024){Bhargava}, {Russell}, {Ng}, {Balasubramanian}, {Zhang}, {Ravi}, {Jadoliya}, {Bhattacharyya}, {Pahari}, {Homan}, {Marshall}, {Chakrabarty}, {Carotenuto}, \& {Kaushik}}]{2024arXiv241100350B}
{Bhargava}, Y., {Russell}, T.~D., {Ng}, M., {et~al.} 2024, arXiv e-prints, arXiv:2411.00350, \dodoi{10.48550/arXiv.2411.00350}

\bibitem[{{Blandford} \& {Payne}(1982)}]{1982MNRAS.199..883B}
{Blandford}, R.~D., \& {Payne}, D.~G. 1982, \mnras, 199, 883, \dodoi{10.1093/mnras/199.4.883}

\bibitem[{{Bobrikova} {et~al.}(2024{\natexlab{a}}){Bobrikova}, {Di Marco}, {La Monaca}, {Poutanen}, {Forsblom}, \& {Loktev}}]{2024A&A...688A.217B}
{Bobrikova}, A., {Di Marco}, A., {La Monaca}, F., {et~al.} 2024{\natexlab{a}}, \aap, 688, A217, \dodoi{10.1051/0004-6361/202450207}

\bibitem[{{Bobrikova} {et~al.}(2025){Bobrikova}, {Poutanen}, \& {Loktev}}]{2025A&A...696A.181B}
{Bobrikova}, A., {Poutanen}, J., \& {Loktev}, V. 2025, \aap, 696, A181, \dodoi{10.1051/0004-6361/202452358}

\bibitem[{{Bobrikova} {et~al.}(2024{\natexlab{b}}){Bobrikova}, {Forsblom}, {Di Marco}, {La Monaca}, {Poutanen}, {Ng}, {Ravi}, {Loktev}, {Kajava}, {Ursini}, {Veledina}, {Rogantini}, {Salmi}, {Bianchi}, {Capitanio}, {Done}, {Fabiani}, {Gnarini}, {Heyl}, {Kaaret}, {Matt}, {Muleri}, {Nitindala}, {Rankin}, {Weisskopf}, {Agudo}, {Antonelli}, {Bachetti}, {Baldini}, {Baumgartner}, {Bellazzini}, {Bongiorno}, {Bonino}, {Brez}, {Bucciantini}, {Castellano}, {Cavazzuti}, {Chen}, {Ciprini}, {Costa}, {De Rosa}, {Del Monte}, {Di Gesu}, {Di Lalla}, {Donnarumma}, {Doroshenko}, {Dov{\v{c}}iak}, {Ehlert}, {Enoto}, {Evangelista}, {Ferrazzoli}, {Garc{\'\i}a}, {Gunji}, {Hayashida}, {Iwakiri}, {Jorstad}, {Karas}, {Kislat}, {Kitaguchi}, {Kolodziejczak}, {Krawczynski}, {Latronico}, {Liodakis}, {Maldera}, {Manfreda}, {Marin}, {Marinucci}, {Marscher}, {Marshall}, {Massaro}, {Mitsuishi}, {Mizuno}, {Negro}, {Ng}, {O'Dell}, {Omodei}, {Oppedisano}, {Papitto}, {Pavlov}, {Peirson}, {Perri}, {Pesce-Rollins}, {Petrucci}, {Pilia}, {Possenti},
  {Puccetti}, {Ramsey}, {Ratheesh}, {Roberts}, {Romani}, {Sgr{\`o}}, {Slane}, {Soffitta}, {Spandre}, {Swartz}, {Tamagawa}, {Tavecchio}, {Taverna}, {Tawara}, {Tennant}, {Thomas}, {Tombesi}, {Trois}, {Tsygankov}, {Turolla}, {Vink}, {Wu}, {Xie}, \& {Zane}}]{2024A&A...688A.170B}
{Bobrikova}, A., {Forsblom}, S.~V., {Di Marco}, A., {et~al.} 2024{\natexlab{b}}, \aap, 688, A170, \dodoi{10.1051/0004-6361/202449318}

\bibitem[{{CASA Team} {et~al.}(2022){CASA Team}, {Bean}, {Bhatnagar}, {Castro}, {Donovan Meyer}, {Emonts}, {Garcia}, {Garwood}, {Golap}, {Gonzalez Villalba}, {Harris}, {Hayashi}, {Hoskins}, {Hsieh}, {Jagannathan}, {Kawasaki}, {Keimpema}, {Kettenis}, {Lopez}, {Marvil}, {Masters}, {McNichols}, {Mehringer}, {Miel}, {Moellenbrock}, {Montesino}, {Nakazato}, {Ott}, {Petry}, {Pokorny}, {Raba}, {Rau}, {Schiebel}, {Schweighart}, {Sekhar}, {Shimada}, {Small}, {Steeb}, {Sugimoto}, {Suoranta}, {Tsutsumi}, {van Bemmel}, {Verkouter}, {Wells}, {Xiong}, {Szomoru}, {Griffith}, {Glendenning}, \& {Kern}}]{2022PASP..134k4501C}
{CASA Team}, {Bean}, B., {Bhatnagar}, S., {et~al.} 2022, \pasp, 134, 114501, \dodoi{10.1088/1538-3873/ac9642}

\bibitem[{{Chandrasekhar}(1960)}]{1960ratr.book.....C}
{Chandrasekhar}, S. 1960, {Radiative transfer}

\bibitem[{{Cocchi} {et~al.}(2023){Cocchi}, {Gnarini}, {Fabiani}, {Ursini}, {Poutanen}, {Capitanio}, {Bobrikova}, {Farinelli}, {Paizis}, {Sidoli}, {Veledina}, {Bianchi}, {Di Marco}, {Ingram}, {Kajava}, {La Monaca}, {Matt}, {Malacaria}, {Miku{\v{s}}incov{\'a}}, {Rankin}, {Zane}, {Agudo}, {Antonelli}, {Bachetti}, {Baldini}, {Baumgartner}, {Bellazzini}, {Bongiorno}, {Bonino}, {Brez}, {Bucciantini}, {Castellano}, {Cavazzuti}, {Chen}, {Ciprini}, {Costa}, {De Rosa}, {Del Monte}, {Di Gesu}, {Di Lalla}, {Donnarumma}, {Doroshenko}, {Dov{\v{c}}iak}, {Ehlert}, {Enoto}, {Evangelista}, {Ferrazzoli}, {Garcia}, {Gunji}, {Hayashida}, {Heyl}, {Iwakiri}, {Jorstad}, {Kaaret}, {Karas}, {Kislat}, {Kitaguchi}, {Kolodziejczak}, {Krawczynski}, {Latronico}, {Liodakis}, {Maldera}, {Manfreda}, {Marin}, {Marinucci}, {Marscher}, {Marshall}, {Massaro}, {Mitsuishi}, {Mizuno}, {Muleri}, {Negro}, {Ng}, {O'Dell}, {Omodei}, {Oppedisano}, {Papitto}, {Pavlov}, {Peirson}, {Perri}, {Pesce-Rollins}, {Petrucci}, {Pilia}, {Possenti}, {Puccetti},
  {Ramsey}, {Ratheesh}, {Roberts}, {Romani}, {Sgr{\`o}}, {Slane}, {Soffitta}, {Spandre}, {Swartz}, {Tamagawa}, {Tavecchio}, {Taverna}, {Tawara}, {Tennant}, {Thomas}, {Tombesi}, {Trois}, {Tsygankov}, {Turolla}, {Vink}, {Weisskopf}, {Wu}, \& {Xie}}]{2023A&A...674L..10C}
{Cocchi}, M., {Gnarini}, A., {Fabiani}, S., {et~al.} 2023, \aap, 674, L10, \dodoi{10.1051/0004-6361/202346275}

\bibitem[{{Corbel} {et~al.}(2000){Corbel}, {Fender}, {Tzioumis}, {Nowak}, {McIntyre}, {Durouchoux}, \& {Sood}}]{Corbel2000A&A...359..251C}
{Corbel}, S., {Fender}, R.~P., {Tzioumis}, A.~K., {et~al.} 2000, \aap, 359, 251, \dodoi{10.48550/arXiv.astro-ph/0003460}

\bibitem[{{Curran} {et~al.}(2014){Curran}, {Coriat}, {Miller-Jones}, {Armstrong}, {Edwards}, {Sivakoff}, {Woudt}, {Altamirano}, {Belloni}, {Corbel}, {Fender}, {K{\"o}rding}, {Krimm}, {Markoff}, {Migliari}, {Russell}, {Stevens}, \& {Tzioumis}}]{Curran2014MNRAS.437.3265C}
{Curran}, P.~A., {Coriat}, M., {Miller-Jones}, J.~C.~A., {et~al.} 2014, \mnras, 437, 3265, \dodoi{10.1093/mnras/stt2125}

\bibitem[{{D'A{\`\i}} {et~al.}(2014){D'A{\`\i}}, {Iaria}, {Di Salvo}, {Riggio}, {Burderi}, \& {Robba}}]{2014A&A...564A..62D}
{D'A{\`\i}}, A., {Iaria}, R., {Di Salvo}, T., {et~al.} 2014, \aap, 564, A62, \dodoi{10.1051/0004-6361/201322044}

\bibitem[{{Di Marco} {et~al.}(2022){Di Marco}, {Costa}, {Muleri}, {Soffitta}, {Fabiani}, {La Monaca}, {Rankin}, {Xie}, {Bachetti}, {Baldini}, {Baumgartner}, {Bellazzini}, {Brez}, {Castellano}, {Del Monte}, {Di Lalla}, {Ferrazzoli}, {Latronico}, {Maldera}, {Manfreda}, {O'Dell}, {Perri}, {Pesce-Rollins}, {Puccetti}, {Ramsey}, {Ratheesh}, {Sgr{\`o}}, {Spandre}, {Tennant}, {Tobia}, {Trois}, \& {Weisskopf}}]{2022AJ....163..170D}
{Di Marco}, A., {Costa}, E., {Muleri}, F., {et~al.} 2022, \aj, 163, 170, \dodoi{10.3847/1538-3881/ac51c9}

\bibitem[{{Di Marco} {et~al.}(2023){Di Marco}, {Soffitta}, {Costa}, {Ferrazzoli}, {La Monaca}, {Rankin}, {Ratheesh}, {Xie}, {Baldini}, {Del Monte}, {Ehlert}, {Fabiani}, {Kim}, {Muleri}, {O'Dell}, {Ramsey}, {Rubini}, {Sgr{\`o}}, {Silvestri}, {Tennant}, \& {Weisskopf}}]{2023AJ....165..143D}
{Di Marco}, A., {Soffitta}, P., {Costa}, E., {et~al.} 2023, \aj, 165, 143, \dodoi{10.3847/1538-3881/acba0f}

\bibitem[{{Di Marco} {et~al.}(2025){Di Marco}, {La Monaca}, {Bobrikova}, {Stella}, {Papitto}, {Poutanen}, {Baglio}, {Bachetti}, {Loktev}, {Pilia}, \& {Rogantini}}]{2025ApJ...979L..47D}
{Di Marco}, A., {La Monaca}, F., {Bobrikova}, A., {et~al.} 2025, \apjl, 979, L47, \dodoi{10.3847/2041-8213/ada7f8}

\bibitem[{{Diaz Trigo} {et~al.}(2010){Diaz Trigo}, {Sidoli}, {Parmar}, \& {Boirin}}]{2010AIPC.1248..153D}
{Diaz Trigo}, M., {Sidoli}, L., {Parmar}, A., \& {Boirin}, L. 2010, in American Institute of Physics Conference Series, Vol. 1248, X-ray Astronomy 2009; Present Status, Multi-Wavelength Approach and Future Perspectives, ed. A.~{Comastri}, L.~{Angelini}, \& M.~{Cappi} (AIP), 153--154, \dodoi{10.1063/1.3475177}

\bibitem[{{Done} {et~al.}(2007){Done}, {Gierli{\'n}ski}, \& {Kubota}}]{2007A&ARv..15....1D}
{Done}, C., {Gierli{\'n}ski}, M., \& {Kubota}, A. 2007, \aapr, 15, 1, \dodoi{10.1007/s00159-007-0006-1}

\bibitem[{{Fabiani} {et~al.}(2024{\natexlab{a}}){Fabiani}, {Capitanio}, {Iaria}, {Poutanen}, {Gnarini}, {Ursini}, {Farinelli}, {Bobrikova}, {Steiner}, {Svoboda}, {Anitra}, {Baglio}, {Carotenuto}, {Del Santo}, {Ferrigno}, {Lewis}, {Russell}, {Russell}, {van den Eijnden}, {Cocchi}, {Di Marco}, {La Monaca}, {Liu}, {Rankin}, {Weisskopf}, {Xie}, {Bianchi}, {Burderi}, {Di Salvo}, {Egron}, {Illiano}, {Kaaret}, {Matt}, {Miku{\v{s}}incov{\'a}}, {Muleri}, {Papitto}, {Agudo}, {Antonelli}, {Bachetti}, {Baldini}, {Baumgartner}, {Bellazzini}, {Bongiorno}, {Bonino}, {Brez}, {Bucciantini}, {Castellano}, {Cavazzuti}, {Chen}, {Ciprini}, {Costa}, {De Rosa}, {Del Monte}, {Di Gesu}, {Di Lalla}, {Donnarumma}, {Doroshenko}, {Dov{\v{c}}iak}, {Ehlert}, {Enoto}, {Evangelista}, {Ferrazzoli}, {Garcia}, {Gunji}, {Hayashida}, {Heyl}, {Iwakiri}, {Jorstad}, {Karas}, {Kislat}, {Kitaguchi}, {Kolodziejczak}, {Krawczynski}, {Latronico}, {Liodakis}, {Maldera}, {Manfreda}, {Marin}, {Marinucci}, {Marscher}, {Marshall}, {Massaro}, {Mitsuishi},
  {Mizuno}, {Negro}, {Ng}, {O'Dell}, {Omodei}, {Oppedisano}, {Pavlov}, {Peirson}, {Perri}, {Pesce-Rollins}, {Petrucci}, {Pilia}, {Possenti}, {Puccetti}, {Ramsey}, {Ratheesh}, {Roberts}, {Romani}, {Sgr{\`o}}, {Slane}, {Soffitta}, {Spandre}, {Swartz}, {Tamagawa}, {Tavecchio}, {Taverna}, {Tawara}, {Tennant}, {Thomas}, {Tombesi}, {Trois}, {Tsygankov}, {Turolla}, {Vink}, {Wu}, \& {Zane}}]{2024AA...684A.137F}
{Fabiani}, S., {Capitanio}, F., {Iaria}, R., {et~al.} 2024{\natexlab{a}}, \aap, 684, A137, \dodoi{10.1051/0004-6361/202347374}

\bibitem[{{Fabiani} {et~al.}(2024{\natexlab{b}}){Fabiani}, {Capitanio}, {Iaria}, {Poutanen}, {Gnarini}, {Ursini}, {Farinelli}, {Bobrikova}, {Steiner}, {Svoboda}, {Anitra}, {Baglio}, {Carotenuto}, {Del Santo}, {Ferrigno}, {Lewis}, {Russell}, {Russell}, {van den Eijnden}, {Cocchi}, {Di Marco}, {La Monaca}, {Liu}, {Rankin}, {Weisskopf}, {Xie}, {Bianchi}, {Burderi}, {Di Salvo}, {Egron}, {Illiano}, {Kaaret}, {Matt}, {Miku{\v{s}}incov{\'a}}, {Muleri}, {Papitto}, {Agudo}, {Antonelli}, {Bachetti}, {Baldini}, {Baumgartner}, {Bellazzini}, {Bongiorno}, {Bonino}, {Brez}, {Bucciantini}, {Castellano}, {Cavazzuti}, {Chen}, {Ciprini}, {Costa}, {De Rosa}, {Del Monte}, {Di Gesu}, {Di Lalla}, {Donnarumma}, {Doroshenko}, {Dov{\v{c}}iak}, {Ehlert}, {Enoto}, {Evangelista}, {Ferrazzoli}, {Garcia}, {Gunji}, {Hayashida}, {Heyl}, {Iwakiri}, {Jorstad}, {Karas}, {Kislat}, {Kitaguchi}, {Kolodziejczak}, {Krawczynski}, {Latronico}, {Liodakis}, {Maldera}, {Manfreda}, {Marin}, {Marinucci}, {Marscher}, {Marshall}, {Massaro}, {Mitsuishi},
  {Mizuno}, {Negro}, {Ng}, {O'Dell}, {Omodei}, {Oppedisano}, {Pavlov}, {Peirson}, {Perri}, {Pesce-Rollins}, {Petrucci}, {Pilia}, {Possenti}, {Puccetti}, {Ramsey}, {Ratheesh}, {Roberts}, {Romani}, {Sgr{\`o}}, {Slane}, {Soffitta}, {Spandre}, {Swartz}, {Tamagawa}, {Tavecchio}, {Taverna}, {Tawara}, {Tennant}, {Thomas}, {Tombesi}, {Trois}, {Tsygankov}, {Turolla}, {Vink}, {Wu}, \& {Zane}}]{2024A&A...684A.137F}
---. 2024{\natexlab{b}}, \aap, 684, A137, \dodoi{10.1051/0004-6361/202347374}

\bibitem[{{Farinelli} {et~al.}(2024){Farinelli}, {Waghmare}, {Ducci}, \& {Santangelo}}]{2024A&A...684A..62F}
{Farinelli}, R., {Waghmare}, A., {Ducci}, L., \& {Santangelo}, A. 2024, \aap, 684, A62, \dodoi{10.1051/0004-6361/202348915}

\bibitem[{{Farinelli} {et~al.}(2023){Farinelli}, {Fabiani}, {Poutanen}, {Ursini}, {Ferrigno}, {Bianchi}, {Cocchi}, {Capitanio}, {De Rosa}, {Gnarini}, {Kislat}, {Matt}, {Mikusincova}, {Muleri}, {Agudo}, {Antonelli}, {Bachetti}, {Baldini}, {Baumgartner}, {Bellazzini}, {Bongiorno}, {Bonino}, {Brez}, {Bucciantini}, {Castellano}, {Cavazzuti}, {Ciprini}, {Costa}, {Del Monte}, {Di Gesu}, {Di Lalla}, {Di Marco}, {Donnarumma}, {Doroshenko}, {Dov{\v{c}}iak}, {Ehlert}, {Enoto}, {Evangelista}, {Ferrazzoli}, {Garcia}, {Gunji}, {Hayashida}, {Heyl}, {Iwakiri}, {Jorstad}, {Karas}, {Kitaguchi}, {Kolodziejczak}, {Krawczynski}, {La Monaca}, {Latronico}, {Liodakis}, {Maldera}, {Manfreda}, {Marin}, {Marscher}, {Marshall}, {Mitsuishi}, {Mizuno}, {Ng}, {O'Dell}, {Omodei}, {Oppedisano}, {Papitto}, {Pavlov}, {Peirson}, {Perri}, {Pesce-Rollins}, {Petrucci}, {Pilia}, {Possenti}, {Puccetti}, {Ramsey}, {Rankin}, {Ratheesh}, {Romani}, {Sgr{\`o}}, {Slane}, {Soffitta}, {Spandre}, {Tamagawa}, {Tavecchio}, {Taverna}, {Tawara}, {Tennant},
  {Thomas}, {Tombesi}, {Trois}, {Tsygankov}, {Turolla}, {Vink}, {Weisskopf}, {Wu}, {Xie}, \& {Zane}}]{2023MNRAS.519.3681F}
{Farinelli}, R., {Fabiani}, S., {Poutanen}, J., {et~al.} 2023, \mnras, 519, 3681, \dodoi{10.1093/mnras/stac3726}

\bibitem[{{Fender}(2006)}]{2006csxs.book..381F}
{Fender}, R. 2006, in Compact stellar X-ray sources, Vol.~39, 381--419, \dodoi{10.48550/arXiv.astro-ph/0303339}

\bibitem[{{Fender} {et~al.}(1999){Fender}, {Garrington}, {McKay}, {Muxlow}, {Pooley}, {Spencer}, {Stirling}, \& {Waltman}}]{Fender1999MNRAS.304..865F}
{Fender}, R.~P., {Garrington}, S.~T., {McKay}, D.~J., {et~al.} 1999, \mnras, 304, 865, \dodoi{10.1046/j.1365-8711.1999.02364.x}

\bibitem[{{Fomalont} {et~al.}(2001){Fomalont}, {Geldzahler}, \& {Bradshaw}}]{Fomalont2001ApJ...558..283F}
{Fomalont}, E.~B., {Geldzahler}, B.~J., \& {Bradshaw}, C.~F. 2001, \apj, 558, 283, \dodoi{10.1086/322479}

\bibitem[{{Fragos} {et~al.}(2010){Fragos}, {Tremmel}, {Rantsiou}, \& {Belczynski}}]{2010ApJ...719L..79F}
{Fragos}, T., {Tremmel}, M., {Rantsiou}, E., \& {Belczynski}, K. 2010, \apjl, 719, L79, \dodoi{10.1088/2041-8205/719/1/L79}

\bibitem[{{Giridharan} {et~al.}(2024){Giridharan}, {Thomas}, {Gudennavar}, \& {Bubbly}}]{2024MNRAS.52711855G}
{Giridharan}, L., {Thomas}, N.~T., {Gudennavar}, S.~B., \& {Bubbly}, S.~G. 2024, \mnras, 527, 11855, \dodoi{10.1093/mnras/stad3941}

\bibitem[{{Gnarini} {et~al.}(2022){Gnarini}, {Ursini}, {Matt}, {Bianchi}, {Capitanio}, {Cocchi}, {Farinelli}, \& {Zhang}}]{2022MNRAS.514.2561G}
{Gnarini}, A., {Ursini}, F., {Matt}, G., {et~al.} 2022, \mnras, 514, 2561, \dodoi{10.1093/mnras/stac1523}

\bibitem[{{Gnarini} {et~al.}(2025){Gnarini}, {Ursini}, {Matt}, {Bianchi}, {Capitanio}, {Cocchi}, {Fabiani}, {Farinelli}, \& {Tarana}}]{2025A&A...699A.230G}
---. 2025, \aap, 699, A230, \dodoi{10.1051/0004-6361/202554573}

\bibitem[{{Grindlay} \& {Seaquist}(1986)}]{1986ApJ...310..172G}
{Grindlay}, J.~E., \& {Seaquist}, E.~R. 1986, \apj, 310, 172, \dodoi{10.1086/164673}

\bibitem[{{Hasinger} \& {van der Klis}(1989)}]{1989A&A...225...79H}
{Hasinger}, G., \& {van der Klis}, M. 1989, \aap, 225, 79

\bibitem[{{Homan} {et~al.}(1998){Homan}, {van der Klis}, {Wijnands}, {Vaughan}, \& {Kuulkers}}]{1998ApJ...499L..41H}
{Homan}, J., {van der Klis}, M., {Wijnands}, R., {Vaughan}, B., \& {Kuulkers}, E. 1998, \apjl, 499, L41, \dodoi{10.1086/311341}

\bibitem[{{Hughes} {et~al.}(2023){Hughes}, {Sivakoff}, {Macpherson}, {Miller-Jones}, {Tetarenko}, {Altamirano}, {Anderson}, {Belloni}, {Heinz}, {Jonker}, {K{\"o}rding}, {Maitra}, {Markoff}, {Migliari}, {Mooley}, {Rupen}, {Russell}, {Russell}, {Sarazin}, {Soria}, \& {Tudose}}]{Hughes2023MNRAS.521..185H}
{Hughes}, A.~K., {Sivakoff}, G.~R., {Macpherson}, C.~E., {et~al.} 2023, \mnras, 521, 185, \dodoi{10.1093/mnras/stad396}

\bibitem[{{Iaria} {et~al.}(2014){Iaria}, {Di Salvo}, {Burderi}, {Riggio}, {D'A{\`\i}}, \& {Robba}}]{2014A&A...561A..99I}
{Iaria}, R., {Di Salvo}, T., {Burderi}, L., {et~al.} 2014, \aap, 561, A99, \dodoi{10.1051/0004-6361/201322328}

\bibitem[{{Kaddouh} {et~al.}(2024){Kaddouh}, {Sudha}, \& {Ludlam}}]{2024RNAAS...8..243K}
{Kaddouh}, M.~A., {Sudha}, M., \& {Ludlam}, R.~M. 2024, Research Notes of the American Astronomical Society, 8, 243, \dodoi{10.3847/2515-5172/ad7e22}

\bibitem[{{Kashyap} {et~al.}(2025){Kashyap}, {Maccarone}, {Ng}, {Pattie}, {Ravi}, \& {Marshall}}]{2025ApJ...986..207K}
{Kashyap}, U., {Maccarone}, T.~J., {Ng}, M., {et~al.} 2025, \apj, 986, 207, \dodoi{10.3847/1538-4357/adda35}

\bibitem[{{Kislat} {et~al.}(2015){Kislat}, {Clark}, {Beilicke}, \& {Krawczynski}}]{2015APh....68...45K}
{Kislat}, F., {Clark}, B., {Beilicke}, M., \& {Krawczynski}, H. 2015, Astroparticle Physics, 68, 45, \dodoi{10.1016/j.astropartphys.2015.02.007}

\bibitem[{{La Monaca} {et~al.}(2024{\natexlab{a}}){La Monaca}, {Di Marco}, {Ludlam}, {Bobrikova}, {Poutanen}, {Li}, \& {Xie}}]{2024A&A...691A.253L}
{La Monaca}, F., {Di Marco}, A., {Ludlam}, R.~M., {et~al.} 2024{\natexlab{a}}, \aap, 691, A253, \dodoi{10.1051/0004-6361/202451966}

\bibitem[{{La Monaca} {et~al.}(2024{\natexlab{b}}){La Monaca}, {Di Marco}, {Poutanen}, {Bachetti}, {Motta}, {Papitto}, {Pilia}, {Xie}, {Bianchi}, {Bobrikova}, {Costa}, {Deng}, {Ge}, {Illiano}, {Jia}, {Krawczynski}, {Lai}, {Liu}, {Mastroserio}, {Muleri}, {Rankin}, {Soffitta}, {Veledina}, {Ambrosino}, {Del Santo}, {Chen}, {Garcia}, {Kaaret}, {Russell}, {Wei}, {Zhang}, {Zuo}, {Arzoumanian}, {Cocchi}, {Gnarini}, {Farinelli}, {Gendreau}, {Ursini}, {Weisskopf}, {Zane}, {Agudo}, {Antonelli}, {Baldini}, {Baumgartner}, {Bellazzini}, {Bongiorno}, {Bonino}, {Brez}, {Bucciantini}, {Capitanio}, {Castellano}, {Cavazzuti}, {Chen}, {Ciprini}, {De Rosa}, {Del Monte}, {Di Gesu}, {Di Lalla}, {Donnarumma}, {Doroshenko}, {Dov{\v{c}}iak}, {Ehlert}, {Enoto}, {Evangelista}, {Fabiani}, {Ferrazzoli}, {Gunji}, {Hayashida}, {Heyl}, {Iwakiri}, {Jorstad}, {Karas}, {Kislat}, {Kitaguchi}, {Kolodziejczak}, {Latronico}, {Liodakis}, {Maldera}, {Manfreda}, {Marin}, {Marinucci}, {Marscher}, {Marshall}, {Massaro}, {Matt}, {Mitsuishi}, {Mizuno},
  {Negro}, {Ng}, {O'Dell}, {Omodei}, {Oppedisano}, {Pavlov}, {Peirson}, {Perri}, {Pesce-Rollins}, {Petrucci}, {Possenti}, {Puccetti}, {Ramsey}, {Ratheesh}, {Roberts}, {Romani}, {Sgr{\`o}}, {Slane}, {Spandre}, {Swartz}, {Tamagawa}, {Tavecchio}, {Taverna}, {Tawara}, {Tennant}, {Thomas}, {Tombesi}, {Trois}, {Tsygankov}, {Turolla}, {Vink}, {Wu}, \& {IXPE Collaboration}}]{2024ApJ...960L..11L}
{La Monaca}, F., {Di Marco}, A., {Poutanen}, J., {et~al.} 2024{\natexlab{b}}, \apjl, 960, L11, \dodoi{10.3847/2041-8213/ad132d}

\bibitem[{{Lin} {et~al.}(2007){Lin}, {Remillard}, \& {Homan}}]{2007ApJ...667.1073L}
{Lin}, D., {Remillard}, R.~A., \& {Homan}, J. 2007, \apj, 667, 1073, \dodoi{10.1086/521181}

\bibitem[{{Madej} {et~al.}(2014){Madej}, {Jonker}, {D{\'\i}az Trigo}, \& {Mi{\v{s}}kovi{\v{c}}ov{\'a}}}]{2014MNRAS.438..145M}
{Madej}, O.~K., {Jonker}, P.~G., {D{\'\i}az Trigo}, M., \& {Mi{\v{s}}kovi{\v{c}}ov{\'a}}, I. 2014, \mnras, 438, 145, \dodoi{10.1093/mnras/stt2119}

\bibitem[{{Mitsuda} {et~al.}(1989){Mitsuda}, {Inoue}, {Nakamura}, \& {Tanaka}}]{1989PASJ...41...97M}
{Mitsuda}, K., {Inoue}, H., {Nakamura}, N., \& {Tanaka}, Y. 1989, \pasj, 41, 97

\bibitem[{{Rankin} {et~al.}(2024){Rankin}, {La Monaca}, {Di Marco}, {Poutanen}, {Bobrikova}, {Kravtsov}, {Muleri}, {Pilia}, {Veledina}, {Fender}, {Kaaret}, {Kim}, {Marinucci}, {Marshall}, {Papitto}, {Tennant}, {Tsygankov}, {Weisskopf}, {Wu}, {Zane}, {Ambrosino}, {Farinelli}, {Gnarini}, {Agudo}, {Antonelli}, {Bachetti}, {Baldini}, {Baumgartner}, {Bellazzini}, {Bianchi}, {Bongiorno}, {Bonino}, {Brez}, {Bucciantini}, {Capitanio}, {Castellano}, {Cavazzuti}, {Chen}, {Ciprini}, {Costa}, {De Rosa}, {Del Monte}, {Di Gesu}, {Di Lalla}, {Donnarumma}, {Doroshenko}, {Dov{\v{c}}iak}, {Ehlert}, {Enoto}, {Evangelista}, {Fabiani}, {Ferrazzoli}, {Garcia}, {Gunji}, {Hayashida}, {Heyl}, {Iwakiri}, {Jorstad}, {Karas}, {Kislat}, {Kitaguchi}, {Kolodziejczak}, {Krawczynski}, {Latronico}, {Liodakis}, {Maldera}, {Manfreda}, {Marin}, {Marscher}, {Massaro}, {Matt}, {Mitsuishi}, {Mizuno}, {Negro}, {Ng}, {O'Dell}, {Omodei}, {Oppedisano}, {Pavlov}, {Peirson}, {Perri}, {Pesce-Rollins}, {Petrucci}, {Possenti}, {Puccetti}, {Ramsey},
  {Ratheesh}, {Roberts}, {Romani}, {Sgr{\`o}}, {Slane}, {Soffitta}, {Spandre}, {Swartz}, {Tamagawa}, {Tavecchio}, {Taverna}, {Tawara}, {Thomas}, {Tombesi}, {Trois}, {Turolla}, {Vink}, \& {Xie}}]{2024ApJ...961L...8R}
{Rankin}, J., {La Monaca}, F., {Di Marco}, A., {et~al.} 2024, \apjl, 961, L8, \dodoi{10.3847/2041-8213/ad1832}

\bibitem[{{Ratheesh} {et~al.}(2024){Ratheesh}, {Dov{\v{c}}iak}, {Krawczynski}, {Podgorn{\'y}}, {Marra}, {Veledina}, {Suleimanov}, {Rodriguez Cavero}, {Steiner}, {Svoboda}, {Marinucci}, {Bianchi}, {Negro}, {Matt}, {Tombesi}, {Poutanen}, {Ingram}, {Taverna}, {West}, {Karas}, {Ursini}, {Soffitta}, {Capitanio}, {Viscolo}, {Manfreda}, {Muleri}, {Parra}, {Beheshtipour}, {Chun}, {Cibrario}, {Di Lalla}, {Fabiani}, {Hu}, {Kaaret}, {Loktev}, {Miku{\v{s}}incov{\'a}}, {Mizuno}, {Omodei}, {Petrucci}, {Puccetti}, {Rankin}, {Zane}, {Zhang}, {Agudo}, {Antonelli}, {Bachetti}, {Baldini}, {Baumgartner}, {Bellazzini}, {Bongiorno}, {Bonino}, {Brez}, {Bucciantini}, {Castellano}, {Cavazzuti}, {Chen}, {Ciprini}, {Costa}, {De Rosa}, {Del Monte}, {Di Gesu}, {Di Marco}, {Donnarumma}, {Doroshenko}, {Ehlert}, {Enoto}, {Evangelista}, {Ferrazzoli}, {Garcia}, {Gunji}, {Hayashida}, {Heyl}, {Iwakiri}, {Jorstad}, {Kislat}, {Kitaguchi}, {Kolodziejczak}, {La Monaca}, {Latronico}, {Liodakis}, {Maldera}, {Marin}, {Marscher}, {Marshall}, {Massaro},
  {Mitsuishi}, {Ng}, {O'Dell}, {Oppedisano}, {Papitto}, {Pavlov}, {Peirson}, {Perri}, {Pesce-Rollins}, {Pilia}, {Possenti}, {Ramsey}, {Roberts}, {Romani}, {Sgr{\`o}}, {Slane}, {Spandre}, {Swartz}, {Tamagawa}, {Tavecchio}, {Tawara}, {Tennant}, {Thomas}, {Trois}, {Tsygankov}, {Turolla}, {Vink}, {Weisskopf}, {Wu}, \& {Xie}}]{2024ApJ...964...77R}
{Ratheesh}, A., {Dov{\v{c}}iak}, M., {Krawczynski}, H., {et~al.} 2024, \apj, 964, 77, \dodoi{10.3847/1538-4357/ad226e}

\bibitem[{{Rawat} {et~al.}(2023){Rawat}, {Garg}, \& {M{\'e}ndez}}]{2023ApJ...949L..43R}
{Rawat}, D., {Garg}, A., \& {M{\'e}ndez}, M. 2023, \apjl, 949, L43, \dodoi{10.3847/2041-8213/acd77b}

\bibitem[{{Russell} {et~al.}(2015){Russell}, {Miller-Jones}, {Curran}, {Soria}, {Altamirano}, {Corbel}, {Coriat}, {Moin}, {Russell}, {Sivakoff}, {Slaven-Blair}, {Belloni}, {Fender}, {Heinz}, {Jonker}, {Krimm}, {K{\"o}rding}, {Maitra}, {Markoff}, {Middleton}, {Migliari}, {Remillard}, {Rupen}, {Sarazin}, {Tetarenko}, {Torres}, {Tudose}, \& {Tzioumis}}]{Russell2015MNRAS.450.1745R}
{Russell}, T.~D., {Miller-Jones}, J.~C.~A., {Curran}, P.~A., {et~al.} 2015, \mnras, 450, 1745, \dodoi{10.1093/mnras/stv723}

\bibitem[{{Saavedra} {et~al.}(2023){Saavedra}, {Garc{\'\i}a}, {Fogantini}, {M{\'e}ndez}, {Combi}, {Luque-Escamilla}, \& {Mart{\'\i}}}]{2023MNRAS.522.3367S}
{Saavedra}, E.~A., {Garc{\'\i}a}, F., {Fogantini}, F.~A., {et~al.} 2023, \mnras, 522, 3367, \dodoi{10.1093/mnras/stad1157}

\bibitem[{{Schnerr} {et~al.}(2003){Schnerr}, {Reerink}, {van der Klis}, {Homan}, {M{\'e}ndez}, {Fender}, \& {Kuulkers}}]{2003A&A...406..221S}
{Schnerr}, R.~S., {Reerink}, T., {van der Klis}, M., {et~al.} 2003, \aap, 406, 221, \dodoi{10.1051/0004-6361:20030682}

\bibitem[{{Stella} \& {Vietri}(1998)}]{1998ApJ...492L..59S}
{Stella}, L., \& {Vietri}, M. 1998, \apjl, 492, L59, \dodoi{10.1086/311075}

\bibitem[{{Sunyaev} \& {Titarchuk}(1985)}]{1985A&A...143..374S}
{Sunyaev}, R.~A., \& {Titarchuk}, L.~G. 1985, \aap, 143, 374

\bibitem[{{Tomaru} {et~al.}(2020){Tomaru}, {Done}, {Ohsuga}, {Odaka}, \& {Takahashi}}]{2020MNRAS.497.4970T}
{Tomaru}, R., {Done}, C., {Ohsuga}, K., {Odaka}, H., \& {Takahashi}, T. 2020, \mnras, 497, 4970, \dodoi{10.1093/mnras/staa2254}

\bibitem[{{Ueda} {et~al.}(2001){Ueda}, {Asai}, {Yamaoka}, {Dotani}, \& {Inoue}}]{2001ApJ...556L..87U}
{Ueda}, Y., {Asai}, K., {Yamaoka}, K., {Dotani}, T., \& {Inoue}, H. 2001, \apjl, 556, L87, \dodoi{10.1086/323007}

\bibitem[{{Ueda} {et~al.}(2004){Ueda}, {Murakami}, {Yamaoka}, {Dotani}, \& {Ebisawa}}]{2004ApJ...609..325U}
{Ueda}, Y., {Murakami}, H., {Yamaoka}, K., {Dotani}, T., \& {Ebisawa}, K. 2004, \apj, 609, 325, \dodoi{10.1086/420973}

\bibitem[{{Ursini} {et~al.}(2024){Ursini}, {Gnarini}, {Capitanio}, {Bobrikova}, {Cocchi}, {Di Marco}, {Fabiani}, {Farinelli}, {La Monaca}, {Rankin}, {Saade}, \& {Poutanen}}]{2024Galax..12...43U}
{Ursini}, F., {Gnarini}, A., {Capitanio}, F., {et~al.} 2024, Galaxies, 12, 43, \dodoi{10.3390/galaxies12040043}

\bibitem[{{van der Klis}(1995)}]{1995foap.conf..213V}
{van der Klis}, M. 1995, in Frontier Objects in Astrophysics and Particle Physics, ed. F.~{Giovannelli} \& G.~{Mannocchi}, 213

\bibitem[{{van der Klis}(2004)}]{2004astro.ph.10551V}
{van der Klis}, M. 2004, arXiv e-prints, astro, \dodoi{10.48550/arXiv.astro-ph/0410551}

\bibitem[{{Wells}(1999)}]{1999JQSRT..62...29W}
{Wells}, R.~J. 1999, \jqsrt, 62, 29, \dodoi{10.1016/S0022-4073(97)00231-8}

\bibitem[{{White} {et~al.}(1988){White}, {Stella}, \& {Parmar}}]{1988ApJ...324..363W}
{White}, N.~E., {Stella}, L., \& {Parmar}, A.~N. 1988, \apj, 324, 363, \dodoi{10.1086/165901}

\bibitem[{{Wilms} {et~al.}(2000){Wilms}, {Allen}, \& {McCray}}]{2000ApJ...542..914W}
{Wilms}, J., {Allen}, A., \& {McCray}, R. 2000, \apj, 542, 914, \dodoi{10.1086/317016}

\end{thebibliography}
\bibliographystyle{aasjournal}



\end{document}